\documentclass[useAMS,usenatbib,usegraphicx]{mn2e}

\usepackage{epsf,rotating}
\usepackage{setspace}
\usepackage{subfigure}
\usepackage{mathrsfs}
\usepackage{amssymb,amsmath}

\newcommand{\lya}{{\rm Ly}\alpha}

\newcommand{\kms}{{\rm km}\,{\rm s}^{-1}}
\newcommand{\cms}{{\rm cm}^{-2}}
\newcommand{\cmc}{{\rm cm}^{-3}}

\newcommand{\Zsolar}{\;{\rm Z}_{\odot}}
\newcommand{\msolar}{{\rm M}_{\odot}}

\newcommand{\CIV}{\hbox{C\,{\sc iv}}}
\newcommand{\CV}{\hbox{C\,{\sc v}}}

\newcommand{\CVII}{\hbox{C\,{\sc vii}}}

\newcommand{\SV}{\hbox{S\,{\sc v}}}

\newcommand{\NIV}{\hbox{N\,{\sc iv}}}
\newcommand{\NV}{\hbox{N\,{\sc v}}}

\newcommand{\OIII}{\hbox{O\,{\sc iii}}}
\newcommand{\OIV}{\hbox{O\,{\sc iv}}}

\newcommand{\OVI}{\hbox{O\,{\sc vi}}}
\newcommand{\OVII}{\hbox{O\,{\sc vii}}}
\newcommand{\OVIII}{\hbox{O\,{\sc viii}}}

\newcommand{\HI}{{\hbox{H\,{\sc i}}}}
\newcommand{\HII}{{\hbox{H\,{\sc ii}}}}

\newcommand{\HeII}{{\hbox{He\,{\sc ii}}}}

\newcommand{\NeV}{{\hbox{Ne\,{\sc v}}}}
\newcommand{\NeVIII}{{\hbox{Ne\,{\sc viii}}}}
\newcommand{\NeIX}{{\hbox{Ne\,{\sc ix}}}}
\newcommand{\MgII}{{\hbox{Mg\,{\sc ii}}}}
\newcommand{\MgX}{{\hbox{Mg\,{\sc x}}}}
\newcommand{\MgXI}{{\hbox{Mg\,{\sc xi}}}}

\newcommand{\nh}{{n_{\rm H}}}

\newcommand{\trec}{\tau_{\rm rec}}
\newcommand{\trecHII}{\tau_{\rm rec,HII}}
\newcommand{\trecCV}{\tau_{\rm rec,CV}}
\newcommand{\trecCVI}{\tau_{\rm rec,CVI}}

\newcommand{\trecOVI}{\tau_{\rm rec,OVI}}
\newcommand{\trecOVII}{\tau_{\rm rec,OVII}}
\newcommand{\trecOVIII}{\tau_{\rm rec,OVIII}}
\newcommand{\trecNeVIII}{\tau_{\rm rec,NeVIII}}
\newcommand{\trecNeIX}{\tau_{\rm rec,NeIX}}

\newcommand{\tAGN}{\tau_{\rm AGN}}
\newcommand{\tionHI}{\tau_{\rm ion,HI}}

\newcommand{\tionOVI}{\tau_{\rm ion,OVI}}

\newcommand{\fnuunits}{{\rm erg}\,{\rm s}^{-1}\,{\rm cm}^{-2}\,{\rm Hz}^{-1}}
\newcommand{\Jnuunits}{{\rm erg}\,{\rm s}^{-1}\,{\rm cm}^{-2}\,{\rm Hz}^{-1}\,{\rm sr}^{-1}}
\newcommand{\ergs}{{\rm erg}\,{\rm s}^{-1}}

\DeclareMathSymbol{\la}{3}{AMSa}{46}
\DeclareMathSymbol{\ga}{3}{AMSa}{38}

\title[AGN Proximity Zone Fossils]{AGN proximity zone fossils and the delayed recombination of metal lines}

\author[B. D. Oppenheimer \& J. Schaye]{
\parbox[t]{\textwidth}{\vspace{-1cm}
Benjamin D. Oppenheimer$^{1,2}$, Joop Schaye$^1$}
\\\\$^1$ Leiden Observatory, Leiden University, PO Box 9513, 2300 RA Leiden, the Netherlands
\\$^2$ CASA, Department of Astrophysical and Planetary Sciences, University of Colorado, Boulder, CO 80309, USA
}

\begin{document}

\pubyear{2013}

\maketitle

\label{firstpage}

\begin{abstract}
We model the time-dependent evolution of metal-enriched intergalactic
and circumgalactic gas exposed to the fluctuating radiation field from
an active galactic nucleus (AGN).  We consider diffuse gas densities
($\nh=10^{-5}-10^{-2.5} \cmc$) exposed to the extra-galactic
background (EGB) and initially in thermal equilibrium ($T\sim
10^4-10^{4.5}$ K).  Once the proximate AGN field turns on, additional
photo-ionisation rapidly ionises the $\HI$ and metals.  The enhanced
AGN radiation field turns off after a typical AGN lifetime
($\tAGN=1-20$ Myr) and the field returns to the EGB intensity, but the
metals remain out of ionisation equilibrium for timescales that can
significantly exceed $\tAGN$.  We define this phase as the AGN
proximity zone ``fossil'' phase and show that high ionisation stages
(e.g.\ $\OVI$, $\NeVIII$, $\MgX$) are in general enhanced, while the
abundances of low ions (e.g.\ $\CIV$, $\OIV$, $\MgII$) are reduced.
In contrast, $\HI$ re-equilibrates rapidly ($\ll\tAGN$) owing to its
low neutral fraction at diffuse densities.  We demonstrate that metal
column densities of intervening gas observed in absorption in quasar
sight lines are significantly affected by delayed recombination for a
wide range of densities, metallicities, AGN strengths, AGN lifetimes,
and AGN duty cycles.  As an example, we show that a fossil zone model
can simultaneously reproduce the observed $\NeVIII$, $\MgII$, $\HI$,
and other metal columns of the $z=0.927$ PG1206+259 absorption system
observed by \citet{tri11} using a single, $T\sim 10^4$ K phase model.
At low redshift even moderate-strength AGN that are off for 90\% of
the time could significantly enhance the high-ion metal columns in the
circum-galactic media of galaxies observed without active AGN. Fossil
proximity zones may be particularly important during the quasar era,
$z \sim 2-5$. Indeed, we demonstrate that at these redshifts a large
fraction of the metal-enriched intergalactic medium may consist of
out-of-equilibrium fossil zones. AGN proximity zone fossils allow a
whole new class of non-equilibrium solutions that may be applicable to
a large fraction of observed metal absorbers and which could
potentially change the inferred physical conditions and masses of
diffuse gases.
\end{abstract}


\section{Introduction}  

Active galactic nuclei (AGN) provide the majority of photons capable
of ionising intergalactic hydrogen over most of the age of the
Universe \citep[e.g.][]{shu99,haa12}.  Some of the most commonly observed
metal absorption lines in the intergalactic medium (IGM), including
$\CIV$ and $\OVI$, have ionisation potentials at extreme UV (EUV)
energies where their main photo-ionising sources are AGN.  The mean
free paths of ionising photons with energies greater than 1 Rydberg
are long enough to warrant the assumption that the ionising
extra-galactic background (EGB) is uniform in its spatial distribution
\citep[e.g.][]{zuo92}.  The main criterion for this assumption is that
the mean free paths of photons are longer than the clustering length
of photo-ionising sources \citep[e.g.][]{fau09}.

This assumption obviously breaks down near sources of ionising
radiation.  Particularly notable are the proximity zones of QSOs,
where the locally generated ionising radiation at EUV energies may
exceed the EGB by orders of magnitude.  The simple nature of hydrogen,
having only two ionisation states, allows an elegant relationship
where, in ionisation equilibrium, the neutral fraction scales in
inverse linear proportion to the ionisation field strength if hydrogen
is photo-ionised.  A dramatic example of this are the proximity zones
of high-$z$ quasars, where the $\lya$ forest absorption is
significantly reduced by the enhanced AGN radiation field as the QSO
is approached \citep[e.g.][]{car87,baj88,sco00,bol07}. QSO pairs can
probe the transverse proximity effect in which the background QSO line
of sight intersects the foreground QSO proximity zone, where the
radiation field is increased by orders of magnitude compared to the
typical EGB \citep[e.g.][]{jak03,schi04,hen06,gon08}.

Metal absorption also reflects the effects of increased ionisation in
quasar proximity zones.  Proximate absorbers are often defined as
being within 5000 $\kms$ of the QSO velocity along the line of sight,
and show some distinct signatures such as $\OVI$ absorption with
little or no $\HI$ absorption \citep[e.g.][]{tri08}.  \citet{gon08}
observe unique metal-line systems at $z\sim2.5$ in transverse
proximate zones with an ionisation enhancement of 10-200$\times$,
where $\OVI$ is enhanced and $\CIV$ is weakened.  These signatures
suggest that enhanced photo-ionisation affects metal lines, but in a
different way than hydrogen.  The multiple ionisation levels of heavy
elements makes the response to increased ionisation more complex:
metal ion fractions of order unity are possible at diffuse IGM
densities (i.e.\ much higher than $\HI$ fractions); plus these
fractions are not a monotonic function of density or temperature like
that of $\HI$, but instead peak at a preferred density or temperature
for an intermediate ionisation state.  In the case of an increased
ionisation field, the peak ionisation fraction for a metal species
will move in linear proportion with the photo-ionisation rate to
higher density if the species is photo-ionised and in equilibrium.

Another difference between hydrogen and metal species is the timescale
that it takes each to achieve ionisation equilibrium when a proximate
AGN turns on and off.  The ionisation timescales, relevant for the AGN
turn on, are longer for metal species with higher ionisation
potentials, because fewer ionising photons are produced by the AGN at
these higher energies.  However, the more important difference is the
longer timescales metals take to reach ionisation equilibrium after
the AGN turns off.  Although most metal species have shorter
recombination timescales, $\trec$, than $\HII$, the relevant timescale
to achieve equilibrium is $\trec$ times the equilibrium ionisation
fraction of the recombined species.  For example, at typical warm
($T\sim 10^4$ K) IGM densities of the $\lya$ forest ($\nh\sim 10^{-4}
\cmc$), $\trecHII \sim 10^9$ yr, but the $\HI$ ionisation fraction,
$f_{\rm HI}$, is $\sim 10^{-4}$; hence the equilibration timescale is
only $\sim 10^5$ yr, because hydrogen only has to recombine to a very
low neutral fraction to achieve equilibrium.  Metal-line species like
$\CIV$ and $\OVI$ have recombination timescales that are $\ga 10^7$ yr
at similar warm IGM densities, but typical ionisation fractions much
closer to unity.  Therefore, timescales to recombine to equilibrium
are also $\ga 10^7$ yr -- much longer than for $\HI$.  Furthermore,
the existence of multiple ionisation states for metals means that
there are a series of ions to recombine through, and hence multiple
recombination timescales, further extending the total equilibration
timescale.

AGN proximity zone fossils are defined as proximity zones in which the
AGN has turned off, but ionisation equilibrium has not yet been
achieved.  The fossil lifetime can last much longer than the AGN-on
phase, which is often calculated to be $\sim 10^6-10^8$ yr
\citep[e.g.][]{hai01,mar01,jak03,schi04,hopk06,gon08}.  In this work,
we use the non-equilibrium code introduced in \citet{opp13a} to follow
the ionisation and temperature evolution of metal-enriched gas
subjected to a variable ionisation field owing to an AGN.  We consider
several situations where the fossil stage lasts as long as, and
sometimes much longer than the AGN phase.  We predict the absorption
line signatures of these fossil zones, and compare them to some recent
observational results.  We focus on diffuse gas with hydrogen numbers
densities $\nh=10^{-5}-10^{-2.5} \cmc$
(i.e.\ $\rho/\langle\rho\rangle\sim(50-15000)\times(1+z)^{-3}$),
initially at the equilibrium temperature where cooling balances
photo-heating, corresponding to warm IGM temperatures, $T\sim
10^{4}-10^{4.5}$ K.

The first case we will consider may be applicable to the case of the
recently discovered strong $\NeVIII$ absorber detected by
\citet{tri11} in the Cosmic Origins Spectrograph (COS) spectrum of
PG1206+459 at $z\sim 0.927$.  This system is one of the most unique
and most puzzling metal absorbers ever observed, containing both high
ions like $\NeVIII$ and low ions like $\MgII$.  We explore
non-equilibrium photo-ionised models with the AGN on and in the fossil
phase.  We bolster our argument with evidence for a post-starburst,
$>L^*$ galaxy, possibly within 100 proper kpc, which shows signatures
of being an AGN.  We use this as a fiducial example to consider
variations in AGN strength, lifetime, and duty cycles to show how
metal lines can respond to variable ionisation conditions.

We next consider AGN proximity zone fossils at the height of QSO
activity at $z=2.5$.  The observations of metal-line absorption in
transverse QSO proximity zones along the line of sight
\citep[e.g.][]{wor06,wor07,gon08} provide our initial motivation, and
we extend the exploration to the case where the proximity zones turn
into fossil zones after the foreground QSOs turn off.  We argue that
unique absorption signatures should occur where $\HI$ re-equilibrates,
but $\OVI$ and $\CIV$ are still out of equilibrium.  We suggest that
during the AGN era a large fraction of metal-line systems could
reside in proximity zone fossils.

The last example we consider concerns massive, spiral galaxies at
$z=0.25$ that may recently have harboured active AGN, such as
Seyferts, but are currently in the less active or off phase.  We
demonstrate that increases of at least a dex are possible for
circumgalactic (CGM) $\OVI$ time-averaged over fossil phases for normal
Seyfert luminosities turning on 1 Myr out of every 10 Myr.  Our
examples show just how sensitive metal lines can be to moderate AGN,
even if they are only on for a small fraction of time.  As nearly all
galaxies are thought to contain supermassive black holes in their
centres, fossil proximity zones may be very important for the
interpretation of QSO absorption line data.  

This paper is organised as follows.  We introduce our non-equilibrium
code and the methods used in this paper in \S\ref{sec:method}.  We
apply these methods to the AGN proximity zones and proximity zone
fossils for the examples mentioned above in \S\ref{sec:application},
and we summarise in \S\ref{sec:summary}.  Throughout we use proper
distances unless noted otherwise.

\section{Method} \label{sec:method}

We use the non-equilibrium solver introduced in \citet{opp13a} to
follow the time-dependent ionisation states of 11 elements (H, He, C,
N, O, Ne, Mg, Si, S, Ca, and Fe).  The solver also calculates
ion-by-ion cooling and photo-heating to follow the time-dependent
non-equilibrium net cooling, defined as cooling minus photo-heating.
The code is intended to be integrated in hydrodynamic codes, and
includes a treatment to sub-cycle the integration of ionisation and
cooling given the input Courant timestep.

In this work, we add the functionality to our method to allow the
addition of photo-ionisation and photo-heating from a time variable
source, which in this case has the spectral energy distribution
(SED) of an AGN.  The AGN photo-ionisation rate (s$^{-1}$) for an
ionisation state $i$ of atomic species $x$ from a radiation source
with specific intensity $f_{\nu}$ (erg s$^{-1}$ cm$^{-2}$ Hz$^{-1}$)
is

\begin{equation} \label{eqn:gamma}
\Gamma_{x_i, \rm AGN} = \int^{\infty}_{\nu_{0,x_i}} \frac{f_{\nu}}{h \nu} \sigma_{x_i}(\nu) d\nu,
\end{equation}

\noindent and the AGN photo-heating rate (erg s$^{-1}$) is

\begin{equation} \label{eqn:eps}
\epsilon_{x_i, \rm AGN} = \int^{\infty}_{\nu_{0,x_i}} \frac{f_{\nu}}{h \nu} \sigma_{i}(\nu) h (\nu-\nu_{0,x_i})d\nu, 
\end{equation}

\noindent where $\nu$ is frequency, $\nu_{0,x_i}$ is the ionisation
frequency, $\sigma_{x_i}(\nu)$ is photo-ionisation cross-section, and
$h$ is the Planck constant.  Since we track 133 ionisation states of
11 elements, we integrate the above equations for the 122
ionisation states with electrons. 

For the EGB, we replace $f_{\nu}$ with the ionising background SED
averaged over the whole sky, $4 \pi J_{\nu}$, in Equations
\ref{eqn:gamma} and \ref{eqn:eps}, where $J_{\nu}$ has units of erg
s$^{-1}$ cm$^{-2}$ Hz$^{-1}$ sr$^{-1}$.  As explained in
\citet{opp13a}, we calculate $\Gamma_{\rm EGB}$ and $\epsilon_{\rm
  EGB}$ at 50 different redshifts between $z=0-9$ using the
\citet[][hereafter HM01]{haa01} EGB and the same $\sigma_{x_i}(\nu)$
used in CLOUDY ver.\ 10.00 last described in \citet{fer98}.  We use
the HM01 EGB as our fiducial background since this results in good
agreement with the observed column density distribution of $\HI$
\citep{dav10, alt11} and metal line strengths and ratios \citep{sch03,
  opp06, agu08, opp09a, opp12a,rah12}.

The time dependent evolution of the particle number density, $n$, for
each ionisation state, $x_i$, is given by

\begin{equation} \label{eqn:ionstate}
\begin{split}
\frac{dn_{x_i}}{dt}& = n_{x_{i+1}} \alpha_{x_{i+1}} n_e + n_{x_{i-1}}(\beta_{x_{i-1}} n_e + 
\Gamma_{x_{i-1,}\rm EGB} \\
& + \Gamma_{x_{i-1},\rm AGN}) - n_{x_i} ((\alpha_{x_i}+\beta_{x_i}) n_e + \Gamma_{x_i,\rm EGB} + \Gamma_{x_i,\rm AGN}),
\end{split}
\end{equation}
\noindent where $n_e$ is the free electron density (cm$^{-3}$),
$\alpha_{x_i}$ is the total recombination rate coefficient (radiative
plus di-electric, cm$^3$ s$^{-1}$), and $\beta_{x_i}$ is the
collisional ionisation rate coefficient (cm$^3$ s$^{-1}$).  The AGN
subscript indicates that the coefficients have been calculated using
our AGN template spectrum explained below.  Charge transfer and Auger
ionisation are also included in our calculations as explained in
\citet{opp13a}, but were omitted from Equation \ref{eqn:ionstate} for
simplicity.  Charge transfer has negligible effects for the species we
explore as this is more important for cases with significant $\HI$
neutral fractions, which are not the focus of this work.  Including
Auger ionisation increases the ionisation by a small amount for the
Lithium-like ions we explore, but the differences are of the order of
0.1-0.2 dex at most, and usually much less \citep{opp13a}.

The net cooling rate per unit volume ($\Lambda_{\rm net}$, erg
cm$^{-3}$ s$^{-1}$), is the difference between cooling and
photo-heating from the EGB and the AGN,
\begin{equation}
\begin{split}
\Lambda_{{\rm net},{x_i}}(T,z,n_{x_i},n_{e})& = \Lambda'_{x_i}(T) n_{x_i} n_{e} - \epsilon_{x_i,\rm EGB}(z) n_{x_i} \\ 
& - \epsilon_{x_i, \rm AGN}(z) n_{x_i}.
\end{split}
\end{equation}
\noindent The units of the cooling efficiency per ion,
$\Lambda'_{x_i}$ are erg cm$^{3}$ s$^{-1}$, where we divided the
cooling rate per volume (erg cm$^{-3}$ s$^{-1}$) for ion $x_{i}$ by
$n_{x_i} n_{e}$ to achieve density independence of collisional
cooling, which allows us to tabulate these rates as a function of
temperature only.

\begin{figure}
\includegraphics[scale=0.7]{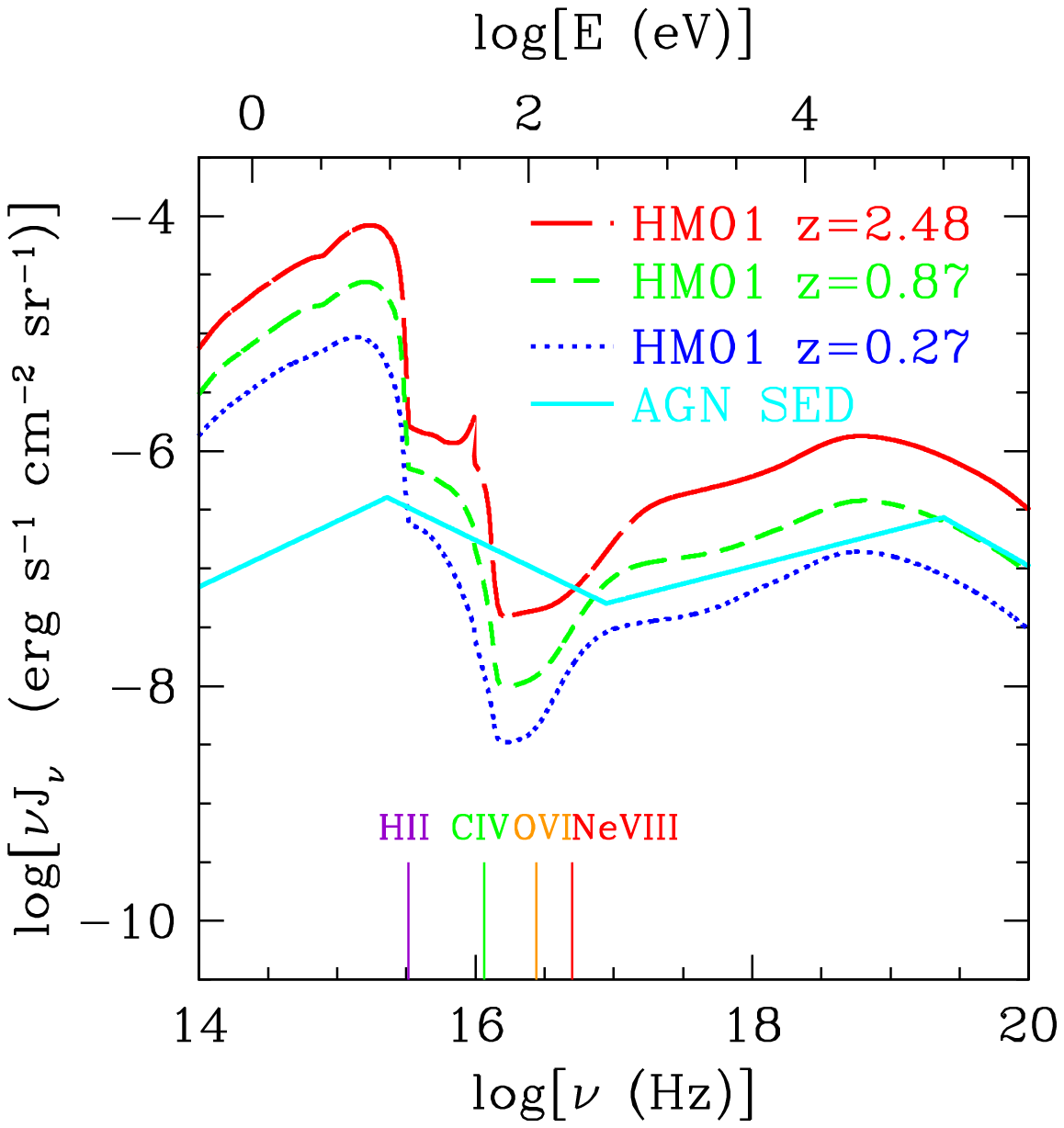}
\caption[]{The \citet{haa01} model for the spectrum of the EGB closest
  to the three redshift regimes ($z=2.5, 0.9$ \& 0.25) we consider,
  along with our AGN template spectrum, normalised to $J_{\rm
    LL}=10^{-22.0} \Jnuunits$.  The EUV power law of the AGN is
  $\nu^{-1.57}$ between 1300 \AA~and 354 eV based on observations by
  \citet{tel02}.  The X-ray portion of the spectrum is assumed to
  follow the CLOUDY AGN template ($\nu^{-0.7}$ until 100 keV and then
  $\nu^{-1.67}$ at higher energies).  The four coloured vertical lines
  at the bottom indicate the ionisation potentials needed to reach the
  corresponding ion listed.  }
\label{fig:SED}
\end{figure}

Figure \ref{fig:SED} compares our AGN template spectrum, $f_{\nu}$ to
the HM01 EGB, $J_{\nu}(z)$ at the three different redshifts ($z=2.48$,
0.87, and 0.27) nearest the three redshifts we consider in this paper
($z=2.5$, 0.9, and 0.25).  The HM01 EGB strength at the Lyman limit at
these three redshifts is from high to low-$z$ log[$J_{\rm LL}$
  ($\Jnuunits$)]$=-21.30, -21.64$, and $-22.16$.

In cyan is the AGN SED normalised to $f_{\rm LL}=10^{-20.9}
\fnuunits$.  Since this is a point source, we can convert its flux
into $J_{\rm LL}$ equivalent units by dividing by $4\pi$ steradians to
get $J_{\rm LL,equ}=10^{-22.0} \Jnuunits$, which has the same units as
the EGB.  Hence, $J_{\rm LL, equ}$ is the intensity the EGB would have
if it were equally bright as the AGN at the Lyman limit.  The spectrum
has an EUV power law with slope $\nu^{-1.57}$ between 1300 \AA~and 354
eV based on observations by \citet{tel02}.  At higher energies, the
spectrum follows $\nu^{-0.7}$ until 100 keV and then $\nu^{-1.67}$ at
higher energies, which follows the template CLOUDY AGN spectrum.  The
most important slope for our consideration is the EUV (i.e.\ 10-354
eV) slope, because these energies correspond to the ionisation
potentials of the metal species we consider and which are marked in
Figure \ref{fig:SED}.  However, the X-ray portion of the spectrum is
also important for two reasons: i) photo-heating rates of species like
$\OVII$ and $\OVIII$ depend on the shape of the X-ray spectrum, and
ii) we convert to observed luminosities of AGN based on their 0.5-2.0
keV luminosities ($L_{0.5-2.0}$).

To convert these units to AGN luminosities, the above flux ($f_{\rm
  LL}=10^{-20.9} \fnuunits$) corresponds to $L_{0.5-2.0} = 10^{42.1}
\ergs$ AGN at 100 proper kpc, which is the luminosity one may expect
from a local Seyfert.  The Seyfert would outshine the $z=0.25$ HM01
EGB at 912 \AA~by 45\% at 100 kpc, but because of the harder AGN
spectrum at the $\OVI$ photo-ionisation edge, 114 eV, the AGN would
outshine the HM01 spectrum by a factor of $23\times$, assuming no
absorption of the AGN spectrum.  In general, for our template AGN
spectrum the conversion from AGN soft X-ray luminosity to equivalent
EGB intensity units is
\begin{equation} 
\begin{split}
J_{\rm LL,equ}= \frac{f_{\rm LL}}{4\pi} = 
10^{-22} \left( \frac{r_{\rm AGN}}{100\ {\rm kpc}}\right)^{-2} \left( \frac{L_{0.5-2.0}}{10^{42.1} \ergs} \right) & \\  \Jnuunits,
\end{split}
\end{equation}
where $r_{\rm AGN}$ is the proper distance from the AGN.  

\section{Applications of AGN proximity zone fossils} \label{sec:application}

We consider three situations spanning a wide range of redshifts, from
$z=2.5$ to $0.25$, that are applicable to recent observations.  In all
cases we assume the gas is being irradiated by the HM01 EGB for the
given redshift, and that it initially has the equilibrium temperature
for which photo-heating balances cooling.  This temperature
corresponds to $T\sim 10^{3.80-4.68}$ K assuming solar abundances for
the densities we consider, $\nh=10^{-5.0}-10^{-2.5} \cmc$.  The main
reasoning behind our parameter choices is that simulations indicate
that a significant fraction of diffuse metals reside along a locus in
density-temperature phase space where photo-heating balances cooling
\citep[e.g.][]{opp06, wie10, wie11}, and that many commonly observed
metal-line species arise from gas at these densities and temperatures
\citep[e.g.][]{opp12a}.  Although solar abundances are used, we will
argue that our theoretical metal columns scale nearly linearly with
metallicity.

In \S\ref{sec:app1} we discuss our first application: an AGN turning
on at $z=0.9$, which is motivated by observations of the very strong
$\NeVIII$ system observed by \citet{tri11} toward PG1206+459, which
appears to reside near a galaxy showing signs of AGN activity.  We pay
particular attention to this example, because of the large number of
metal-line species observed in this system and also because $\NeVIII$
is one of the species most affected by AGN ionisation.  We vary
several parameters including metallicity, density, AGN strength, and
AGN lifetime.  In \S\ref{sec:app2}, we consider the effects on $\CIV$,
$\NV$, and $\OVI$ columns at $z=2.5$, both when a quasar is on and
after it turns off.  We estimate the fraction of the volume of the IGM
altered by fossil zones, and argue that a large fraction of
observed metal lines may be affected.  Lastly, in \S\ref{sec:seyfert},
we explore short duty cycles of $z=0.25$ AGN that significantly
enhance $\OVI$ columns, even while the AGN is off.  This case could be
applicable to normal looking spirals that were recently Seyferts,
where residual non-equilibrium effects of AGN ionisation enhance metal
columns.

The general behaviour that we find can be summarised as follows.  The
enhanced photo-ionisation from the AGN rapidly raises the ionisation
states of metals, thus allowing high-ionisation species to exist at
higher densities than they would in equilibrium with the normal EGB.
After the AGN turns off, the long recombination times of metals, along
with the multi-ionisation state nature of metals and their significant
ionisation fractions (i.e.\ $ f_{x_i} \ga 0.1$ for multiple $i$), mean
that metal columns often only return to equilibrium on a timescale
that is longer than the AGN-on phase.  Thus, a whole new range of
ionisation models that have not been considered are allowed.

\subsection{AGN fossil zones at $z=0.9$}\label{sec:app1}

In our fiducial example, we consider the effects of an enhanced
ionisation in the proximity zone around an AGN that is on for 20 Myr
at $z=0.9$.  This case is directly applicable to observations by COS
at a redshift where a large array of both low- and high-ionisation
species can be observed.  This example is motivated by the extremely
strong $\NeVIII$ ($N_{\rm NeVIII}\sim 10^{15} \cms$) absorber observed
by \citet{tri11} at $z=0.927$ along the PG1206+459 sight line with
well-aligned lower ionisation species.  A foreground galaxy, labelled
$177\_9$ by that work, found at the same redshift at an impact
parameter of 68 kpc, shows signs of being a post-starburst galaxy with
a $B$-band absolute magnitude of $-22$ and even AGN signatures from
[$\NeV$] emission that likely cannot be explained by star formation
alone.

\begin{figure*}
\includegraphics[scale=0.75]{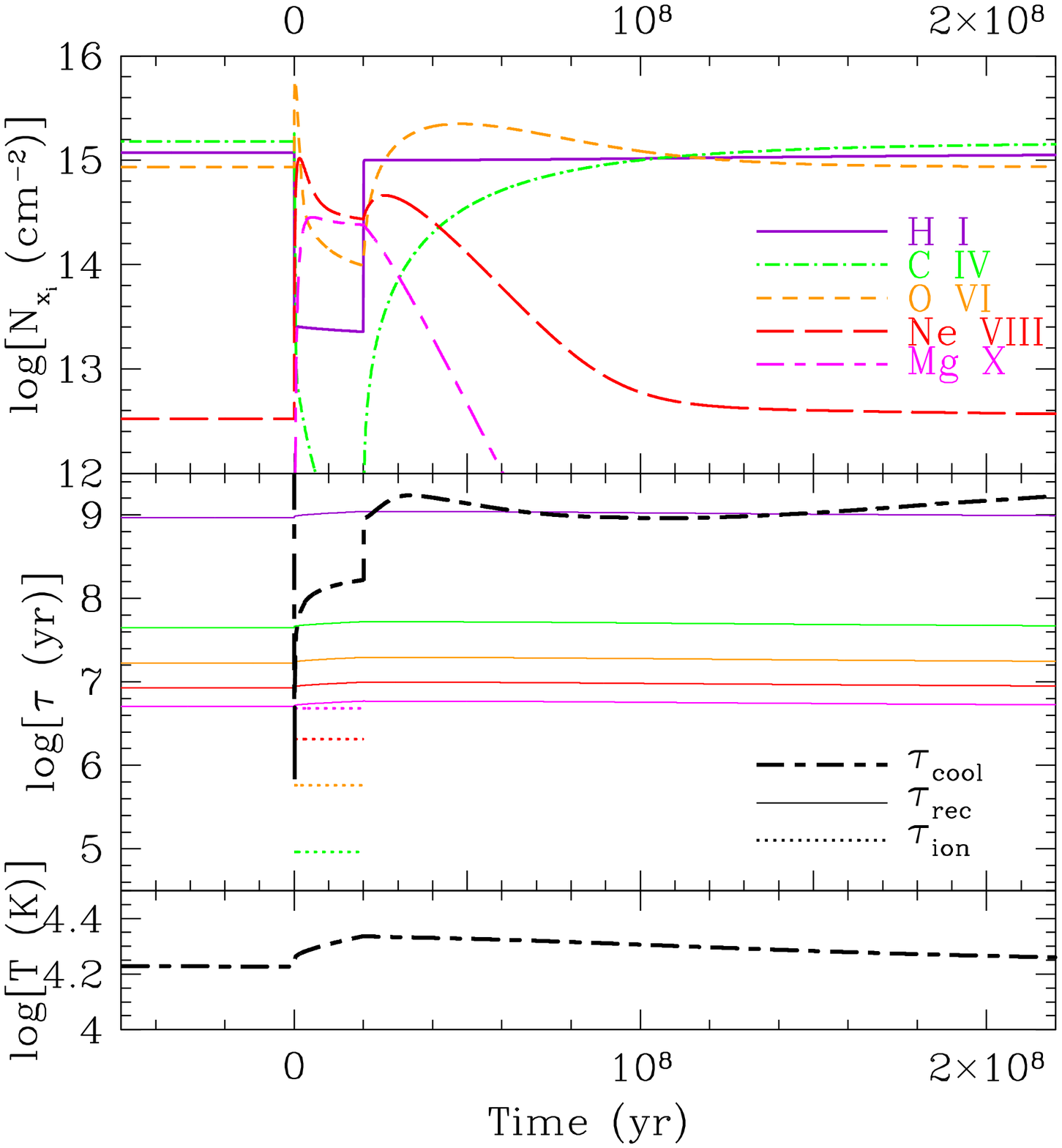}
\caption[]{The evolution of a $\nh = 10^{-4} \cmc$, 74 kpc, solar
  abundance absorber irradiated by the $z=0.9$ HM01 EGB when an AGN
  with intensity $J_{\rm LL,equ}=10^{-20} \Jnuunits$
  ($L_{0.5-2.0}=10^{44.1} \ergs$ at 100 kpc distance) turns on at
  $t=0$ for 20 Myr.  After 20 Myr the AGN turns off but the long
  recombination times of Helium-like metal ions ($\CV$, $\OVII$,
  $\NeIX$, \& $\MgXI$; between 5-50 Myr; middle panel, solid thin
  lines) cause these observable Lithium-like ions to take longer than
  the AGN-on time to reach their equilibrium levels.  Ionisation
  timescales of the Lithium-like ions (middle panel, thick dotted
  lines shown only when the AGN is on) are shorter than 20 Myr when
  the AGN is on.  Photo-heating, primarily of metals, results in a
  slightly increased temperature of the gas that cools on a timescale
  of 1 Gyr (middle panel, thick black dashed lines).  Unlike metals
  and despite a 1 Gyr recombination timescale, hydrogen rapidly
  recombines to equilibrium when the AGN turns off, owing to its very
  small equilibrium neutral fraction.  The columns scale linearly with
  the assumed absorber length and metallicity.}
\label{fig:z09_tevol}
\end{figure*}

Figure \ref{fig:z09_tevol} shows the time evolution of gas enriched to
solar metallicity with $\nh=10^{-4} \cmc$, which corresponds to an
overdensity of 77 at $z=0.9$ (assuming $\Omega_b=0.046$ and a hydrogen
mass fraction of 0.75).  This choice represents a typical density of
metal-enriched CGM gas.  The gas is assumed to lie at its equilibrium
temperature, $T=10^{4.23}$ K, when we mimic the instantaneous turn-on
of the AGN by adding our template AGN spectrum to the HM01 EGB at that
redshift.  We assume an AGN flux of $f_{\rm LL} = 10^{-18.9}
\fnuunits$, which has the $\HI$ ionisation equivalence of a uniform
EGB with $J_{\rm LL,equ}= 10^{-20.0} \Jnuunits$ integrated over $4
\pi$ steradians; this corresponds to $44\times$ the EGB value and a
total ionisation parameter log[$U$]$=0.3$, where $U$ is the ratio of
hydrogen ionising photon density over $\nh$.  Using our template
spectrum and integrating over the soft X-ray band (0.5-2.0 keV), we
calculate that this is the flux at 100 kpc from a
$L_{0.5-2.0}=10^{44.11}$ erg s$^{-1}$ AGN, which appears to be a
reasonable assumption given the observed impact parameter of 68 kpc
for galaxy $177\_9$.

In the top panel we show the column densities of several
observationally accessible ions by assuming an absorber length of 74
kpc, which is the predicted size of an $\HI$ absorber at this density
and temperature according to Equation 3 of \citet{sch01}.  All column
densities scale linearly with the assumed absorber size.  As the AGN
turns on at $t=0$, the $\HI$ column density and ionisation fraction
drop nearly instantaneously by a factor of $44$ as $\tionHI$ is only
1100 yr for this field.\footnote{Note that the ionisation timescale is
  shorter than the light travel times from the AGN to the absorber,
  which we do not consider.}  The ionisation timescales for the metal
species, shown in the middle panel of Figure \ref{fig:z09_tevol}
during the AGN-on phase (thick dotted lines) are longer, $0.1-5$ Myr,
but still less than the assumed AGN lifetime of 20 Myr.  During the
AGN-on phase, the species $\OVI$, $\NeVIII$, and $\MgX$ first rise to
a peak and then fall as they are first ionised from lower ions to
these states and then to higher ions, owing to the large increase in
the radiation field.  On the other hand, $\CIV$, which initially
comprises 27\% of carbon, decreases dramatically as carbon is ionised
all the way to $\CVII$ in under 10 Myr.

The column densities in the AGN-on phase are applicable to proximate
quasar absorption line (QAL) systems observed near the background
source quasar, or to intervening QAL systems where the local ionising
field is significantly enhanced by foreground AGN near the sight line
to the background source quasar.  Absorption systems with
log[$U$]$=0.3$ will have $\NeVIII$ column densities that are
$80-300\times$ stronger than for the HM01 field where log[$U$]$=-1.3$
at $\nh=10^{-4} \cmc$, while $N_{\rm HI}$ will be reduced by a factor
of $50$, and $N_{\rm CIV}$ will be reduced even more.  $N_{\rm OVI}$
can either increase or decrease depending on the AGN lifetime.  $\MgX$
is the most enhanced photo-ionised species shown, increasing more than
$10^4 \times$ in strength to $10^{14.4} \cms$.

The temperature increases by $29\%$ to $T=10^{4.33}$ K due to
photo-heating by the strongly increased ionising field (bottom panel).
The temperature increase results primarily from the ejection of highly
energetic electrons by X-ray photons leading to high-ionisation metal
species (e.g.\ $\CV$, $\OVII$, and $\OVIII$).  For this solar-enriched
case, metal photo-heating dominates over that from $\HeII$, which has
a lower ionisation potential than metal ions.  The photo-heating of
$\HI$ does not contribute to the temperature increase, because in
ionisation equilibrium the increase in the radiation field is exactly
compensated by the corresponding decrease in the neutral fraction, and
the latter is too small to result in significant non-equilibrium
heating.

The AGN proximity zone fossil phase begins after the AGN turns off and
the ionisation field switches back to the uniform HM01 EGB.  The solid
thin lines in the middle panel indicate the recombination timescales
of the various species.  Although $\trecHII\sim 10^9$ yr, $N_{\rm HI}$
reaches equilibrium in a much smaller timescale, $\sim 10^{4.7}$ yr,
because the equilibrium neutral fraction it has to return to is only
$f_{\rm HI} = 10^{-4.3}$.  In contrast, fossil ionised bubbles around
dead quasars can persist for $\sim\trecHII$ during reionisation since
the equilibrium state is neutral \citep{fur08}; however, we are
considering only cases where an EGB ionises the IGM to a high level.
$N_{\rm HI}$ remains slightly smaller than before $t=0$, because
photo-heating increased the temperature of the gas, thereby reducing
the recombination rate.  The cooling timescale of $\sim 1$ Gyr means
that the gas will take of order this time to approach the original
equilibrium temperature and $\HI$ fraction.

Even though the Helium-like metal ion recombination timescales (solid
lines in the middle panel of Figure \ref{fig:z09_tevol}) are
$\sim20-180\times$ shorter than for $\HII$, their recombination lags
are $100-900\times$ longer than for $\HI$, because the equilibrium ion
fractions are a significant fraction of unity.  A case in point is the
timescale of $\sim 50$ Myr that it takes $\CIV$ columns to approach
their previous equilibrium values.  A series of recombination
timescales of species between $\CVII$ and $\CV$ determine the total
time it takes $\CIV$ to return to equilibrium, since carbon is
primarily in $\CVII$ when the AGN turns off. However, the longest is
$\trecCV \sim 40$ Myr, shown as the green solid line in the middle
panel.

$\OVI$ is significantly enhanced during $\sim 70$ Myr after the AGN
turns off, even though its column density is decreased during the
AGN-on phase.  $\OVI$ achieves at least twice as large columns between
33-74 Myr, as oxygen recombines from a majority of $\OVIII$ at $t=20$
Myr to $\OIV$ at $t=95$ Myr.  $\trecOVII=20$ Myr is the bottleneck
slowing the recombination sequence (orange solid line in the middle
panel).

The most dramatic residual observable change is for $\NeVIII$, for
which columns increase 100 fold for the first 15 Myr of the fossil
phase and remain at least $10\times$ stronger for 50 Myr.  Despite the
main recombination bottleneck being $\trecNeIX=10$ Myr, or about half
as long as $\trecOVI$, $\NeVIII$ takes several recombination times to
approach its very low equilibrium value.  Thus $N_{\rm NeVIII} \sim
10^{14} \cms$, which has been observed and attributed to collisionally
ionised gas at $T\sim10^6$ K \citep{sav05, nar09, nar11, mei13}, could
have an alternative colder origin if nearby galaxies currently or
recently harboured an AGN.  

In passing, we note the recent work of \citet{muz13}, who find
extremely strong high ionisation lines (e.g. $\NeVIII$, $\MgX$) in
associated QSO absorbers, which are possibly photo-ionised by the AGN;
however, we note that those absorbers are probably located much nearer
to the supermassive black hole than the extended proximity zones that
are the focus of the present study.  However, the same potential
non-equilibrium fossil effects would apply, albeit on a shorter
timescale given that the densities are likely much higher.  Our method
presented here is general enough to be applied to short-term QSO
fluctuations that keep these zones over-ionised relative to
equilibrium solutions.

\subsubsection{Parameter variations} \label{sec:physparamvar}

We have considered the very specific example of an AGN proximity zone
fossil applicable to enriched gas at densities typical of the
circumgalactic medium.  While our choices appear reasonable, we wish
to broaden the range of physical parameters and situations by varying
the metallicity, density, AGN strength, and AGN lifetime.  We will
relate these variations to situations that may occur in the real
Universe and will argue that fossil zones could significantly change
the interpretation of metal QAL observations.  We will continue to
assume that the metals are cool ($T\sim 10^4-10^{4.5}$ K), residing at
the equilibrium temperature where photo-heating balances radiative
cooling.


\noindent{\bf Metallicity:} Choosing solar metallicity for an
absorption system is not unprecedented as detailed analysis of
observed systems often yield solutions with solar or super-solar
metallicities \citep{per06,pro06,sch07,tri11,tum11}.  However, many of
these systems are attributed to lower density gas, which may on
average be enriched to lower levels if metallicity increases with
density as observations and simulations indicate
\citep[e.g.][]{sch03,opp06,opp09a,wie10,wie11}.  Lowering the
metallicity does not alter the recombination timescales or behaviour
significantly.  The equilibrium temperature in the HM01 ionisation
field increases from $T=10^{4.23}$ K for $Z=\Zsolar$ to $10^{4.49}$ K
for $0.1 \Zsolar$, owing to reduced metal-line cooling, and this
results in only a 0.2 dex reduction of $N_{\rm HI}$.  Hence,
metal-line strengths as a function of time can, to first order, be
scaled with $Z$.  However, if intergalactic metals are poorly mixed on
small scales as some observations and simulations indicate
\citep{sim06a, sch07, opp09a, tep11}, then high metallicities may even
be appropriate at low densities.

\noindent{\bf Density:} Changing the density alters the
non-equilibrium behaviour because the recombination times scale as
$\nh^{-1}$.  The ionisation parameter also changes when assuming a
constant radiation field, meaning that in equilibrium atomic species
are photo-ionised to higher ion states at lower densities.  We explore
here four densities: $\nh=10^{-5.0}$, $10^{-4.5}$, $10^{-4.0}$, and
$10^{-3.5} \cmc$ corresponding to $z=0.90$ overdensities
$\rho/\langle\rho\rangle=7.7$, 24, 77, and 243 respectively.  We
assume an absorber length set by the \citet{sch01} relation ($l=0.74
\nh^{-0.5} \sqrt{T/2\times 10^4}$ kpc), which equals the Jeans scale
in a gas of uniform density and temperature.  We set $T=2\times 10^4$
K in the previous equation for simplicity, even though temperatures
vary.  This size scale, which ranges from 230 to 42 kpc for our
densities, has been shown to reproduce the observed sizes of the
$\lya$ absorbers \citep[e.g][]{sch01, pro04,leh07} and agrees very
well with predictions from hydrodynamic simulations
\citep{dav10,tep12,rah12}.

\begin{figure}
\includegraphics[scale=0.7]{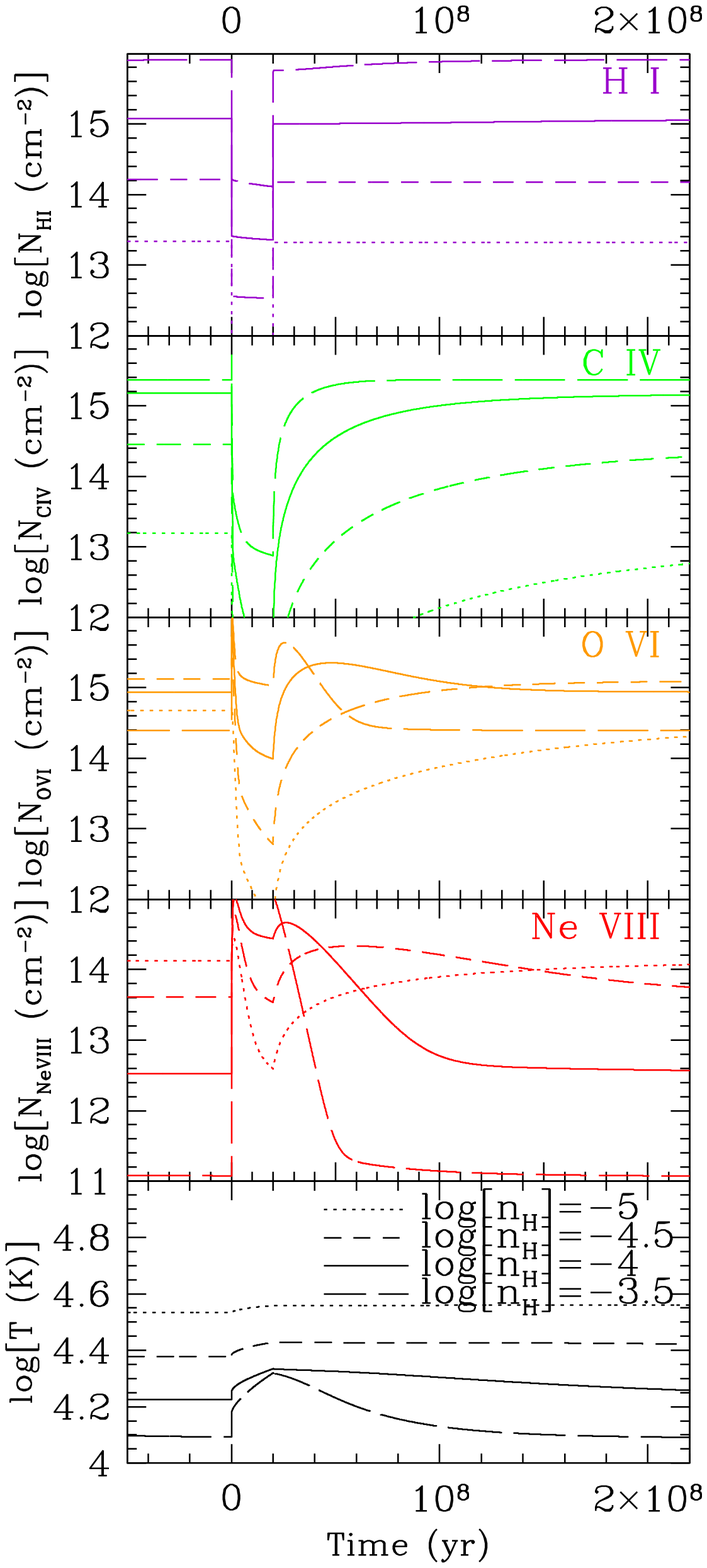}
\caption[]{The density dependence of the non-equilibrium proximity
  effect at $z=0.9$ using an AGN with $J_{\rm LL,equ}=10^{-20.0}
  \Jnuunits$ turning on at $t=0$ for 20 Myr.  The assumed absorber
  lengths decrease from 164 to 29 kpc as density increases from
  $\nh=10^{-5.0}$ to $10^{-3.5} \cmc$.  $N_{\rm HI}$, $N_{\rm CIV}$,
  $N_{\rm OVI}$, $N_{\rm NeVIII}$, and temperature evolution are shown
  from top to bottom.  Ionisation timescales remain unaltered between
  the four cases, but recombination times scale as $\nh^{-1}$ meaning
  that the fossil effect will last longer at lower density.
  Metallicity is assumed to be solar in all cases, but metal columns
  scale with metallicity with the minor exception that photo-heating
  will be lower at lower metallicity and temperatures will not vary as
  much.}
\label{fig:z09_dens}
\end{figure}

Figure \ref{fig:z09_dens}, top panel, shows that the $N_{\rm HI}$ of
the absorbers initially range from $10^{13.4}\rightarrow 10^{15.9}
\cms$ with increasing density, which correspond to observed $\lya$
absorbers that typically exhibit detectable metal-line absorption
\citep{tho08b,tri08,til12}.  As the AGN radiation field turns on and
off, $N_{\rm HI}$ reduces and increases by a factor $50$ on timescales
that are too short to appear in the figure.  A slight deviation occurs
for higher densities where photo-heating raises the temperature during
the AGN more, owing to the greater number of bound electrons available
to ionise from metals.  Although lower densities have a smaller
photo-heating temperature increases (bottom panel), the longer cooling
times of several Gyr mean that a 20 Myr AGN phase has longer lasting
effects on the temperature of the IGM.

The third panel in Figure \ref{fig:z09_dens} shows $N_{\rm OVI}$ for
the four densities.  The overdensity range we explore covers nearly
the same range that simulations predict observed $\OVI$ absorbers to
arise from at $z<0.5$ \citep[e.g.][]{opp09a,tep11,cen11,smi11,opp12a}.
When the AGN turns on, the ionisation timescales are independent of
$\nh$, but the new ionisation equilibrium does depend on $\nh$.  For
$\nh\leq 10^{-4} \cmc$ oxygen is ionised from $\OVI$ to higher states
reducing these columns, while for $\nh\geq 10^{-3.5} \cmc$ lower
ionisation oxygen is ionised to $\OVI$ increasing the columns.  When
the AGN turns off, lower densities take longer to return to ionisation
equilibrium because recombination times scale inversely with $\nh$,
Thus, while the lowest density will have deficient $\OVI$ relative to
the equilibrium case for $>100$ Myr, the highest overdensity will have
enhanced $\OVI$ for a shorter timescale, similar to the AGN lifetime
of 20 Myr in this case.

In contrast to $\OVI$, $N_{\rm CIV}$ is reduced by the AGN for all
densities while $N_{\rm NeVIII}$ shows an increase for $\nh\geq
10^{-4.5}\cmc$ (2nd \& 4th panels of Figure \ref{fig:z09_dens}
respectively).  The $\CIV$ ionisation fraction peaks at $\nh=10^{-3.9}
\cmc$ when photo-ionised by the $z=0.9$ HM01 field at $T\sim 10^4$ K.
Turning on the AGN shifts this peak to $\nh=10^{-1.3} \cmc$.  Hence,
$\CIV$ will be reduced by the AGN, and the lowest densities will take
the longest to recombine from higher ions in the fossil zone.
Conversely, the ionisation fraction of $\NeVIII$ peaks at
$\nh=10^{-5.5} \cmc$ at $T\sim 10^4$ K, which shifts to $10^{-3.3}
\cmc$ with the AGN.  Hence, gas with $\nh=10^{-3.5} \cmc$ can have
$N_{\rm NeVIII}$ columns as high as $10^{15} \cms$ during the AGN
phase, 12,000$\times$ larger than for the HM01 field in equilibrium.
The recombination times of a few Myr means that this strong $\NeVIII$
will be shorter lived in the fossil zone.

\begin{figure}
\includegraphics[scale=0.7]{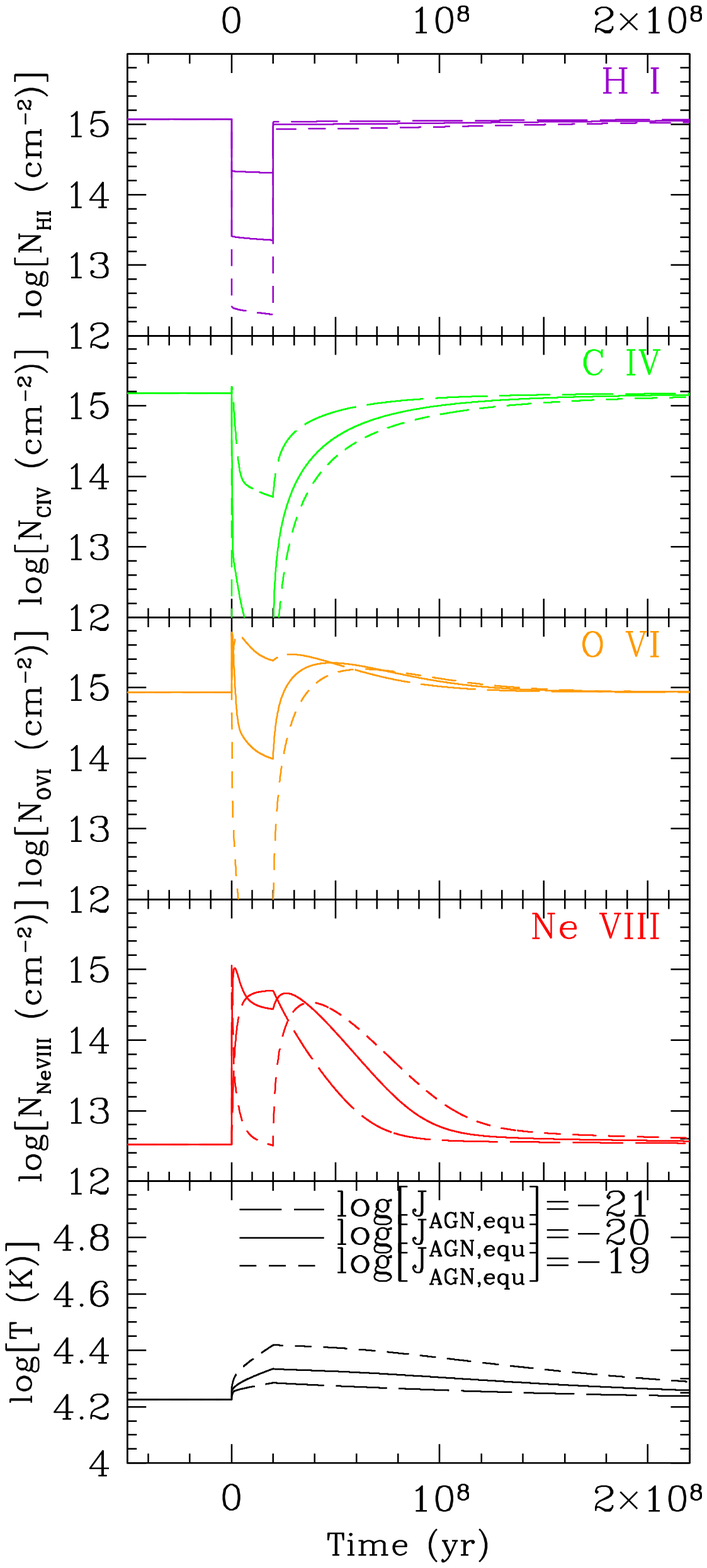}
\caption[]{The dependence of the non-equilibrium proximity effect on
  AGN strength at $z=0.9$ with $\nh=10^{-4} \cmc$ and $l=74$ kpc for
  an AGN turning on at $t=0$ for 20 Myr.  The AGN flux enhancement
  relative to the HM01 EGB ranges from a factor of 4.4 (long dashed
  lines) to 440 (short dashed lines).  The stronger the AGN flux, the
  higher the ionisation state that the metals reach, and the more
  delayed the recombination to equilibrium becomes owing to the need
  to recombine through more ionisation states with significant
  recombination times.  Although the metallicity is assumed to be
  solar, the metal columns scale with metallicity, except for the
  minor effect of photo-heating being less efficient at lower
  metallicity.}
\label{fig:z09_strength}
\end{figure}

\noindent{\bf AGN radiation field increase:} In Figure
\ref{fig:z09_strength}, we consider field increases, $J_{\rm LL,equ}$
of factors 4.4 and 440 to simulate the AGN proximity effect at 3.16
and $0.32\times$ the fiducial radius (i.e.\ 316 and 32 kpc)
respectively for the same strength AGN.  Or alternatively, an AGN 0.1
and $10\times$ as strong at 100 kpc.  We again use our fiducial
density $\nh=10^{-4.0} \cmc$.  The ionisation timescales decrease in
proportion to the field increase, and stronger radiation fields cause
metals to become more ionised in a shorter amount of time.  However,
post-AGN metal-line strengths show qualitatively similar evolution,
because the recombination timescales are the same.  There is a greater
delay for stronger fields, reflecting the additional recombination
time from even higher ionisation states (e.g.\ $\trecCVI$,
$\trecOVIII$).  Even for $J_{\rm LL,equ}=10^{-21} \Jnuunits$, which
corresponds only to a factor $4.4\times$ increase relative to the EGB,
significant enhancements are seen, especially in $\NeVIII$.  The field
increase is insufficient to ionise to $\NeIX$, but post-AGN $\NeVIII$
columns are enhanced 100$\times$ and decline slowly according to
$\trecNeVIII$.  Considering that the $10\times$ smaller field increase
applies to $10^{3/2}\sim 32\times$ more volume, the weaker field case
could be most relevant to intervening QAL, which provide a
volume-weighted sampling of the IGM.

\begin{figure}
\includegraphics[scale=0.7]{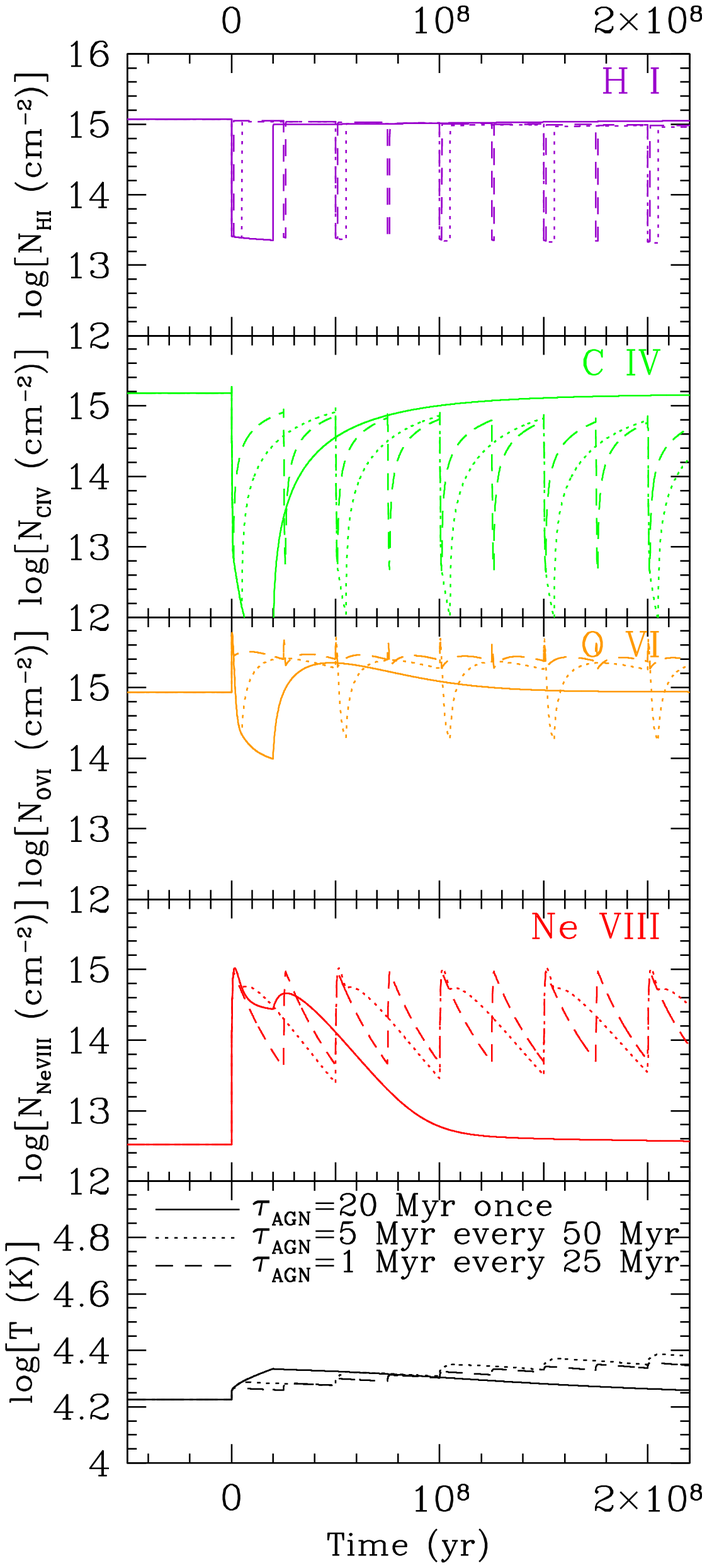}
\caption[]{Different AGN lifetime and duty cycle variations of the
  non-equilibrium proximity effect at at $z=0.9$ with $\nh=10^{-4}
  \cmc$ and $l=52$ kpc for an AGN flux of $J_{\rm LL,equ}=10^{-20}
  \Jnuunits$.  We compare our fiducial case of a single AGN-on phase
  of 20 Myr with an AGN-on phase of 5 Myr every 50 Myr and 1 Myr every
  25 Myr.  The first AGN on phase always starts at $t=0$.  Significant
  non-equilibrium behaviour occurs in all cases significantly
  altering the time-averaged column densities during the AGN-off
  phase.  Photo-heating in bursts can boost average gas temperatures
  as well.  Although the metallicity is assumed to be solar, the metal
  columns scale with metallicity, except for the minor effect of
  photo-heating being less efficient at lower metallicity.}
\label{fig:z09_life}
\end{figure}

\noindent{\bf AGN lifetime:} Thus far we have assumed a single
$\tAGN=20$ Myr lifetime, but AGN lifetimes are predicted to be as
short as $10^6$ yr.  Figure \ref{fig:z09_life} shows what happens if
an AGN is on for shorter periods at multiple times.  We choose to show
an AGN that is on 5 out of every 50 Myr yielding a duty cycle,
$d=10\%$, and an AGN that is on 1 out of every 25 Myr, yielding
$d=4\%$.  For metals, the effect of a shorter lifetime is similar as
that of a weaker AGN: they are not as highly ionised by the AGN, but
the post-AGN recombination times are the same.  A shorter $\tAGN$
results in metals not reaching as highly ionised states.  Post-AGN,
the metals are less likely to have to recombine through these upper
states reducing the number of recombination times to return to
equilibrium.  Nevertheless, the most significant bottleneck is the
longest metal $\trec$ from the Helium-like state ($\CV$, $\OVII$,
$\NeIX$) to the Lithium-like ions we plot.  Hence metal-line evolution
is often similar in these three cases although shifted in time
depending on the strength and of course the frequency of AGN-on
phases.

Even short AGN lifetimes can efficiently enhance metal columns if the
duty cycle is similar to or smaller than the recombination times.  A
case in point is $\NeVIII$ for $\tAGN=1$ Myr in the 4th panel, whose
column density is enhanced by $60\times$ averaged over 0-200 Myr,
despite the AGN being on only 4\% of the time.  For $\tAGN= 1$ Myr,
$\OVI$ is enhanced by a factor of three and remains relatively
constant, owing to a coincidence: $\tionOVI\sim 1$ Myr is similar to
$\tAGN$, and $\trecOVII\sim20$ Myr is similar to the duty cycle
timescale of 25 Myr.  The implications for small duty cycles with
resonances that amplify the fossil effect will be explored further in
\S\ref{sec:seyfert}.  Finally, the bottom panel shows how duty cycles
can pump temperatures up, owing to the combination of short
photo-heating timescales and long cooling timescales if the gas has
solar metallicity.

Multiple AGN-on phases are applicable to the real Universe, although
it is unlikely that an AGN will have such rhythmic repetitions as our
examples here.  Instead, an AGN phase may last on the order of $10^8$
yrs, but fluctuate significantly.  We suggest a model for galaxy
$177\_9$ in \S\ref{sec:PG1206} where it was much brighter in the
recent past, perhaps even a bright QSO, while today it still is a much
weaker AGN.  The point of the duty cycle examples is to show the
fossil effect is non-linear, where multiple short AGN phases have a
much greater effect than one long AGN phase.

\noindent{\bf Extra-galactic background and AGN EUV power laws:} We
have chosen an AGN spectrum with an EUV power law of $\nu^{-1.57}$
based on \citet{tel02}.  The local AGN ionises diffuse gas that is
additionally ionised by the HM01 spectrum, which itself assumes a QSO
EUV input of $\nu^{-1.80}$.  When we apply either the \citet{haa96} or
\citet[][hereafter HM12]{haa12} EGBs that use hard QSO input spectra
of $\nu^{-1.50}$ and $\nu^{-1.57}$ respectively, then the increase in
the radiation field at the ionisation potentials of high ions like
$\OVI$, $\NeVIII$, and $\MgX$ is more muted, but the timescales for
recombination are essentially the same.

Our fiducial example in Figure \ref{fig:z09_tevol} has the following
columns in equilibrium with the HM12 EGB: $N_{\rm HI}=10^{15.3} \cms$
(0.2 dex increase relative to HM01), $N_{\rm CIV}=10^{14.8} \cms$ (0.4
dex decrease), $N_{\rm OVI}=10^{15.5} \cms$ (0.6 dex increase),
$N_{\rm NeVIII}=10^{13.3} \cms$ (0.8 dex increase), and $N_{\rm
  MgX}=10^{10.8} \cms$ (0.4 dex increase).  The addition of the AGN
field with $J_{\rm LL,equ}= 10^{-20.0} \Jnuunits$ leads to the same
AGN-on phase columns as in the fiducial example, meaning that relative
to equilibrium with the HM12 EGB, $\HI$ now declines by 100-fold; the
$\CIV$ decrease, $\NeVIII$ increase, and $\MgX$ increase are all less;
and finally $\OVI$ decreases more.  The metal lines re-equilibrate
during the fossil phase on the same recombination timescales, but
$\NeVIII$ is only $15\times$ higher than the HM12 EGB equilibrium in
the first 15 Myr after the AGN turns off, instead of 100-fold increase
relative to the HM01 field.  The observational consequences are the
same, except that the EGB equilibrium columns for $\NeVIII$ are more
likely to be observable, and fewer recombination timescales are
required to reach EGB equilibrium.  Hence, the biggest difference in
using HM12 or any other background is that photo-ionised equilibrium
columns are different.

In the real Universe, there exists an extremely large range of
observed AGN EUV power laws.  \citet{tel02} and \citet{shu12} together
find a range of $\nu^{-3}$ to $\nu^{0}$ encompassing most of their
measured AGN, and even an outlier with $\nu^{0.56}$, which is the
well-known quasar HE2347-4342 at $z=2.9$ that has been used as a probe
of $\HeII$ reionisation \citep{shu10}.  Hence, the fossil effect may
be even greater for harder AGN, which are more capable of ionising
high ions, while softer AGN will have much less of an effect.
Interestingly, \citet{sco04} provide a sample showing an
anti-correlation between EUV hardness and AGN luminosity.  Their
composite spectrum has a harder EUV slope, $\nu^{-0.56}$, than
\citet{tel02} and \citet{shu12}, which suggests greater fossil
effects.  In fact, they find that their lowest luminosity AGN have EUV
slopes exceeding 0.0 in the low-redshift Universe, which makes our
exploration of weak AGN in \S\ref{sec:seyfert} possibly even more
relevant.  Hence, our choice of a harder AGN EUV spectrum relative to
the EGB for our fiducial example may be quite applicable and could
even underestimate the effect.

\subsubsection{Direct modelling of the $z=0.927$ PG1206+459 $\NeVIII$ absorber} \label{sec:PG1206}

\begin{figure*}
\includegraphics[scale=0.9]{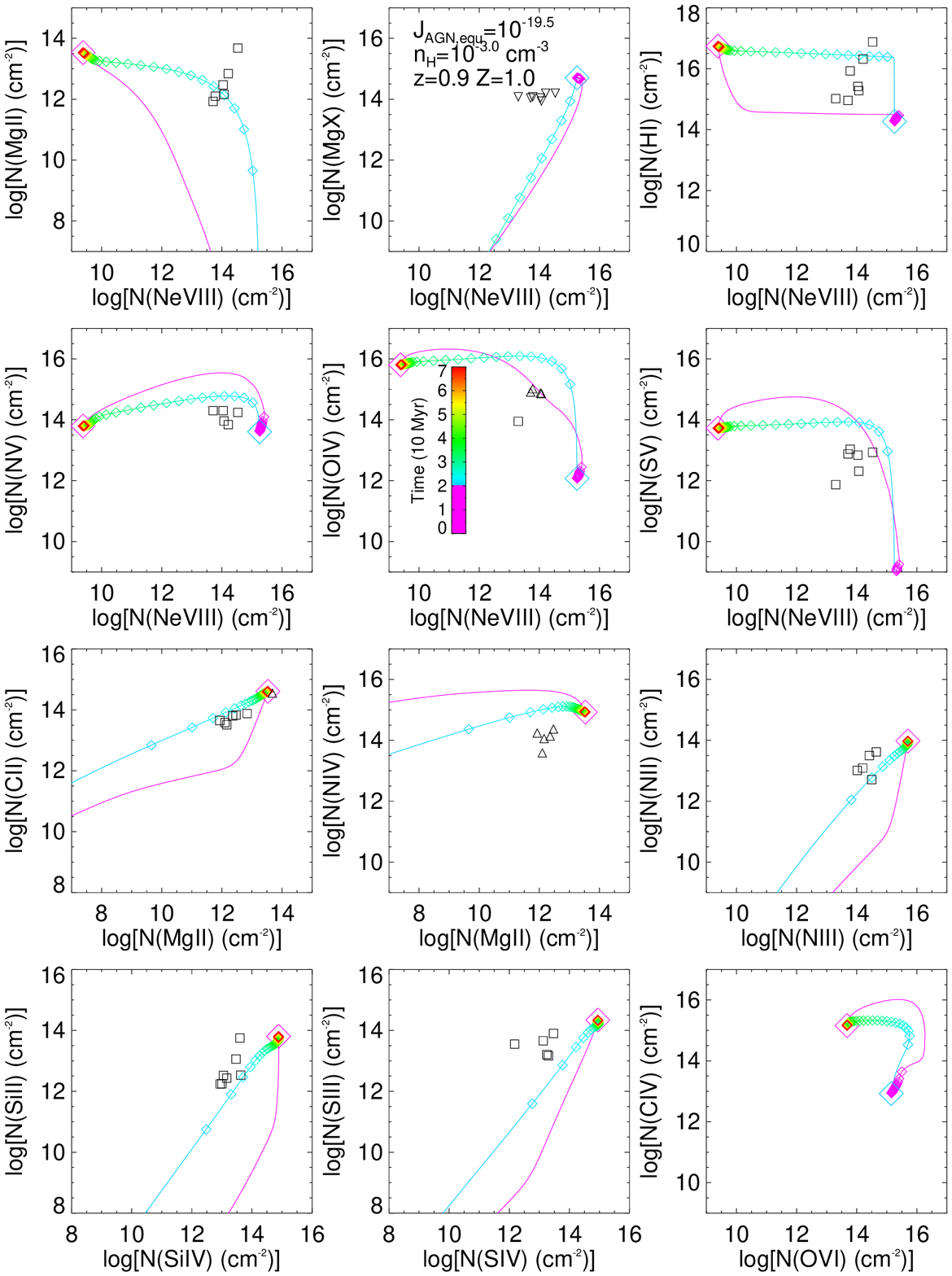}
\caption[]{Evolving column densities of species in the PG1206+459
  $z=0.927$ absorber observed by \citet{tri11} (black squares are
  observed detections, triangles lower limits, upside down triangles
  upper limits) compared to the evolution of column densities
  predicted for a gas with $\nh=10^{-3} \cmc$, $T=10^{3.95}$ K,
  $Z=\Zsolar$, and $l=24$ kpc, irradiated by an AGN with $J_{\rm
    LL,equ}=10^{-19.5} \Jnuunits$ cm$^{-2}$ turning on at $t=0$ and
  off at $t=20$ Myr after which the radiation field returns to the
  constantly present $z=0.9$ HM01 EGB.  Time $t=0$ is shown as a large
  magenta diamond, and $t=20$ Myr as a large blue diamond that
  indicates when the fossil phase begins.  For each of the 12 ion
  pairs, we plot small diamonds every Myr, colour-coded by age in the
  fossil phase (the colour bar is shown in the 5th panel).  The model
  provides a surprisingly good single phase fossil solution 4-5 Myr
  after the AGN turns off for the 16 species considered by
  \citet{tri11} ($\OIII$ observations also agree, but are not shown).
  We make predictions for $\CIV$ and $\OVI$ in the bottom right
  panel.}
\label{fig:z09_ratio}
\end{figure*}

We apply our model to the $z=0.927$ PG1206+459 absorber in Figure
\ref{fig:z09_ratio}, where we plot the evolution of the column
densities of metal-line species and $\HI$ observed by \citet{tri11}.
We explore a range of densities, AGN flux increases, and
metallicities, finding that the model with $\nh=10^{-3.0} \cmc$,
$J_{\rm LL,equ}=10^{-19.5} \Jnuunits$, and $Z=\Zsolar$ provides a
surprisingly good fit for the ions observed by \citet{tri11} using a
single gas phase.  We plot the evolution of one ion against another
ion starting at $t=0$ when the AGN turns on (large magenta diamond),
through the 20 Myr AGN-on phase (magenta), and the fossil phase
(beginning with a large blue diamond at 20 Myr and then changing
colours over the next 50 Myr according to the colour bar).  Diamonds
are plotted every Myr.  The open black squares show the \citet{tri11}
component absorbers, indicating lower limits for saturated absorbers
as triangles in the case of $\OIV$ and $\NIV$, and upper limits as
upside-down triangles for non-detections of $\MgX$.  We combine their
$+65$ and $+108 \kms$ components into o motivated by the blended
$\NeVIII$ detection.

The top left panel shows that we can reproduce the observed $\NeVIII$
and $\MgII$ columns as arising from the same phase 4-5 Myr after the
AGN turns off, which is how we constrained our parameter choice.  No
equilibrium solution with the HM01 background can reproduce both
$\NeVIII$ and $\MgII$, which are shown by \citet{for13} to arise from
very different densities and temperatures in simulations of the CGM
that assume ionisation equilibrium.  The AGN-on phase makes $\NeVIII$
columns large enough, but $\MgII$ is orders of magnitude too weak.
The fossil phase allows a solution for both species that persists in
this case for $\sim 10^6$ yrs given the spacing of the diamonds.

Considering other ions in Figure \ref{fig:z09_ratio}, our solution
agrees with the $\MgX$ upper limits (top middle panel), and because
$\HI$ re-equilibrates to the HM01 solution very rapidly we are able to
reproduce the high neutral hydrogen columns.  Neither of these ions
are reproduced during the AGN-on phase.  Several other observed ions
are plotted against $\NeVIII$ in the second row: $\NV$ is somewhat
high, but could be rectified by not assuming solar abundances as we do
here, $\OIV$ lower limits agree, and $\SV$ does appear to be
over-abundant by a dex.  The bottom six panels show mostly lower ions,
either plotted against $\MgII$ or two ions of the same atomic species
plotted against each other.  The agreement with \citet{tri11} is
usually quite good at $\sim 5$ Myr into the fossil phase, although
nitrogen and silicon may have lower than solar abundances, and sulphur
appears to be greatly over-abundant, by as much as a dex.  The only
species not shown from \citet{tri11} is $\OIII$, for which our model
reproduces the lower limits.  We plot $\OVI$ and $\CIV$ as well
(bottom right panel), even though these species have not been observed
at high resolution, making column density determinations not yet
possible; however \citet{din03} observations indicate very saturated
profiles, which agree with our predictions that $N_{\rm OVI}\sim
10^{15.5} \cms$ and $N_{\rm CIV}\sim 10^{15.0} \cms$.

Figure \ref{fig:z09_ratio} shows merely one possible time-dependent
history for an AGN ionising cool, metal-enriched gas.  Our fiducial
example in Figure \ref{fig:z09_tevol} can also fit $\NeVIII$ and $\NV$
using $10^{-4} \cmc$ gas between 20-50 Myr after the AGN turns off,
but requires a separate phase for the low ions as $\MgII$ is
1000$\times$ too weak.  Figure \ref{fig:z09_ratio} shows a model with
a 140-fold increase in flux at the Lyman limit, but models with 0.5
dex weaker and stronger AGN show nearly identical tracks.  A proximity
zone fossil can enhance metal lines over a significant path length if
the AGN was strong in the past.  We showed that a wide range of AGN
intensities, as would be expected along a large path length through a
proximity zone, will lead to similar recombination time histories and
hence similarly enhanced columns for high-ionisation species at a
given density.  If a proximity zone is many Mpc in size, then enhanced
metal columns should be observed along a similar path length.  A
fossil zone could be traced by unusually strong high-ionisation
species over a large velocity range as in the case of the $\NeVIII$ in
the \citet{tri11} absorber, which extends over 1500 $\kms$.  This
corresponds to a Hubble flow of over 10 proper Mpc implying an
extended fossil zone and a very strong AGN in the past.

We choose an AGN lifetime of 20 Myr, but given that equilibrium AGN-on
columns are achieved in only 1 Myr (i.e. most of the AGN-on evolution
in Figure \ref{fig:z09_ratio} occurs before the first small magenta
diamond appears), the fossil phase evolution will be similar for a
shorter lived AGN.  Our point is that models with time-dependent AGN
that have either turned off in the past, recently turned on, or
fluctuate are worth considering, especially for an extreme case such
as the PG1206+459 absorber with evidence of a post-starburst AGN
within 100 kpc of the sight line.

In summary, we propose a new time-dependent model for the $z=0.927$
PG1206+459 absorber that relies on enhanced photo-ionisation of $T\sim
10^4$ K gas by an AGN that turned off several Myr in the past.  Unlike
equilibrium solutions, out model can reproduce the columns of
$\NeVIII$, $\MgII$, $\HI$, and other metals using a single phase
solution that lasts $\ga 10$ Myr.  This scenario provides an
alternative to the outflow shell model proposed by \citet{tri11} of a
$T>10^5$ K hot gas component and a velocity-coincident cold gas
component, which may be expected for a propagating shock with
post-shock radiatively cooling metal-enriched gas.  In reality, gas at
a variety of densities, and hence decay times, may contribute to the
absorber.  However, we note that the choice of $10^{-3} \cmc$ is well
motivated by ionisation solutions to $T\sim 10^4$ K clouds observed by
\citet{sto13} at $z\la 0.02$.  Our model requires $2\times 10^8
\msolar$ of gas per component, assuming a path length through the
centre of a spherical $r\sim 12$ kpc, $10^{-3.0} \cmc$ solar-enriched
cloud.  This is $200\times$ less mass per component than the
\citet{tri11} thin shell model for their $T>10^5$ K phase.

\subsection{AGN and proximity zone fossils at the peak of QSO activity at $z=2.5$}\label{sec:app2}

Another application of our code is the transverse proximity effect
(TPE) where a foreground QSO ionises gas along the sight line toward a
background QSO.  Recent work has been performed by \citet{hen06},
\citet{wor06,wor07}, and \citet{gon08} who explored the TPE at $z\sim
2.5$, near the peak of QSO activity.  The TPE can be elusive when
traced by $\HI$ alone owing to complicating effects of the overdense
regions creating more $\HI$ absorption, which goes in the opposite
direction of decreased $\HI$ owing to the QSO ionisation \citep[see
  also][]{schi04}.  For example, \citet{hen07} argue that QSO emission
is anisotropic based on the enhancement of Lyman-limit systems in
transverse sight lines indicating that the QSO radiation does not
effect off-sight line directions as often, even after accounting for
the fact that QSOs may live in overdense regions.

\citet{gon08} analysed metal-line systems in a quasar triplet at
$z\sim 2.5$ and found that the TPE was apparent when $\CIV$ and $\OVI$
aligned with $\HI$ are included in the analysis.  They in fact argued
that unusual $N_{\rm OVI}/N_{\rm HI}$ and $N_{\rm CIV}/N_{\rm HI}$
ratios of components in transverse proximity zones indicate that the
QSO radiation field reaches off-sight line directions, and that the
two foreground QSOs, which are extremely bright and rare, more likely
radiate isotropically.  Their work shows that the non-linear
ionisation properties of metal lines, when used in conjunction with
$\HI$, can help constrain the TPE \citep[see also][]{wor06,wor07}.

\subsubsection{The QSO-on phase}

\begin{figure*}
\subfigure{\setlength{\epsfxsize}{0.45\textwidth}\epsfbox{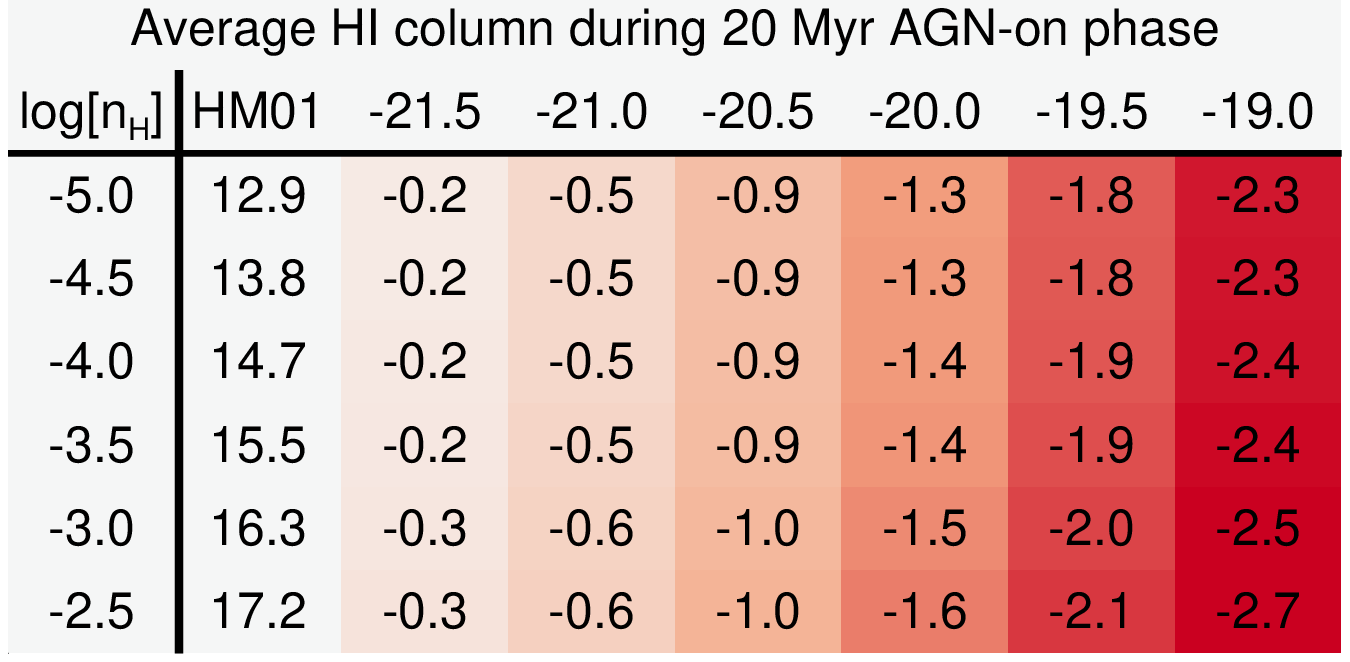}}
\subfigure{\setlength{\epsfxsize}{0.45\textwidth}\epsfbox{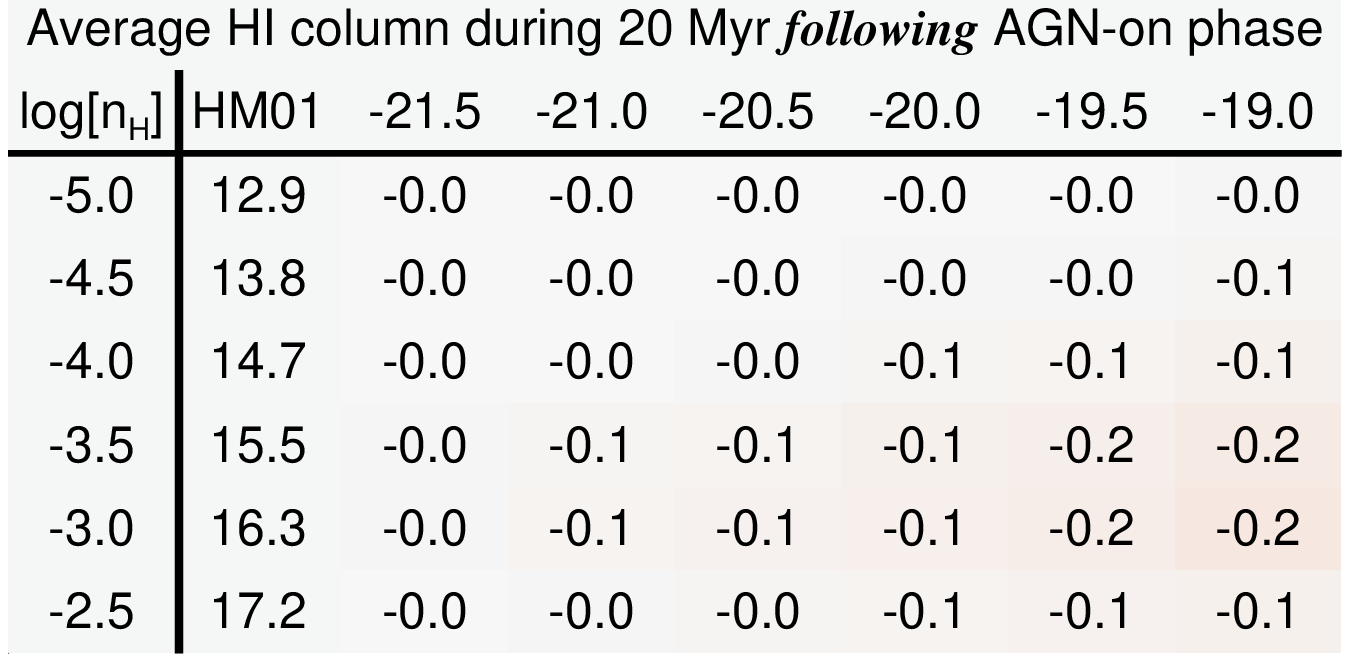}}
\subfigure{\setlength{\epsfxsize}{0.45\textwidth}\epsfbox{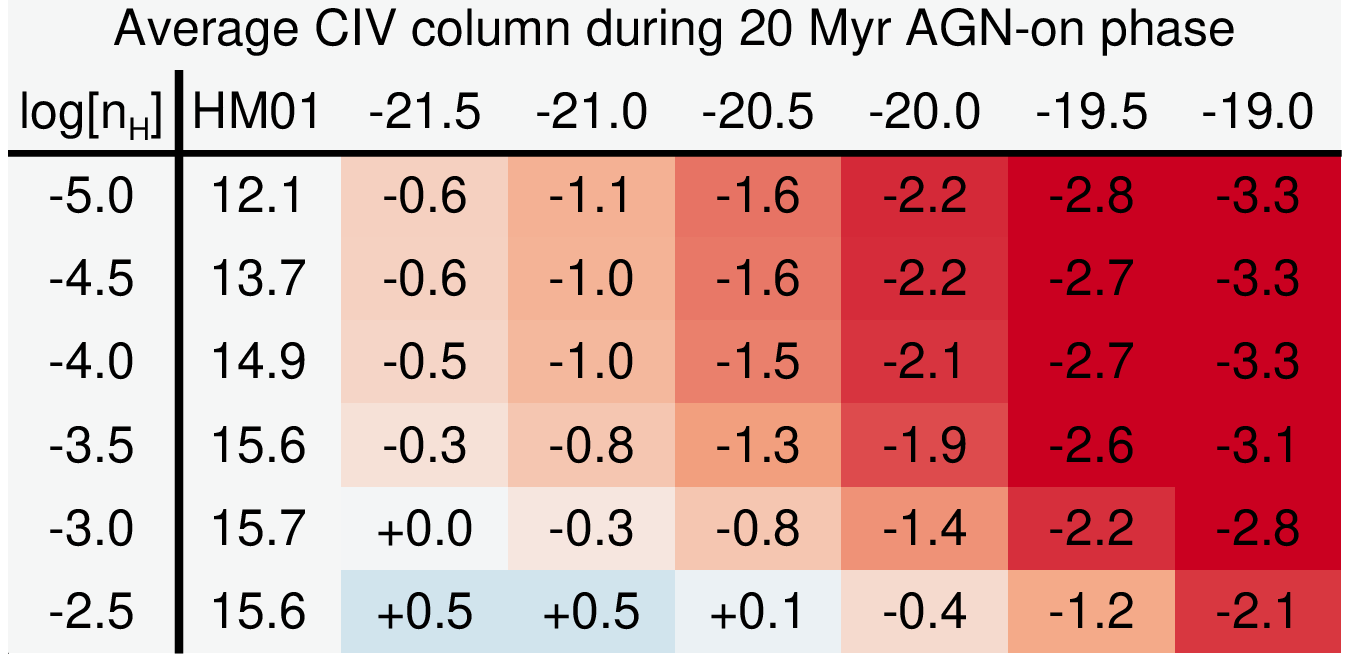}}
\subfigure{\setlength{\epsfxsize}{0.45\textwidth}\epsfbox{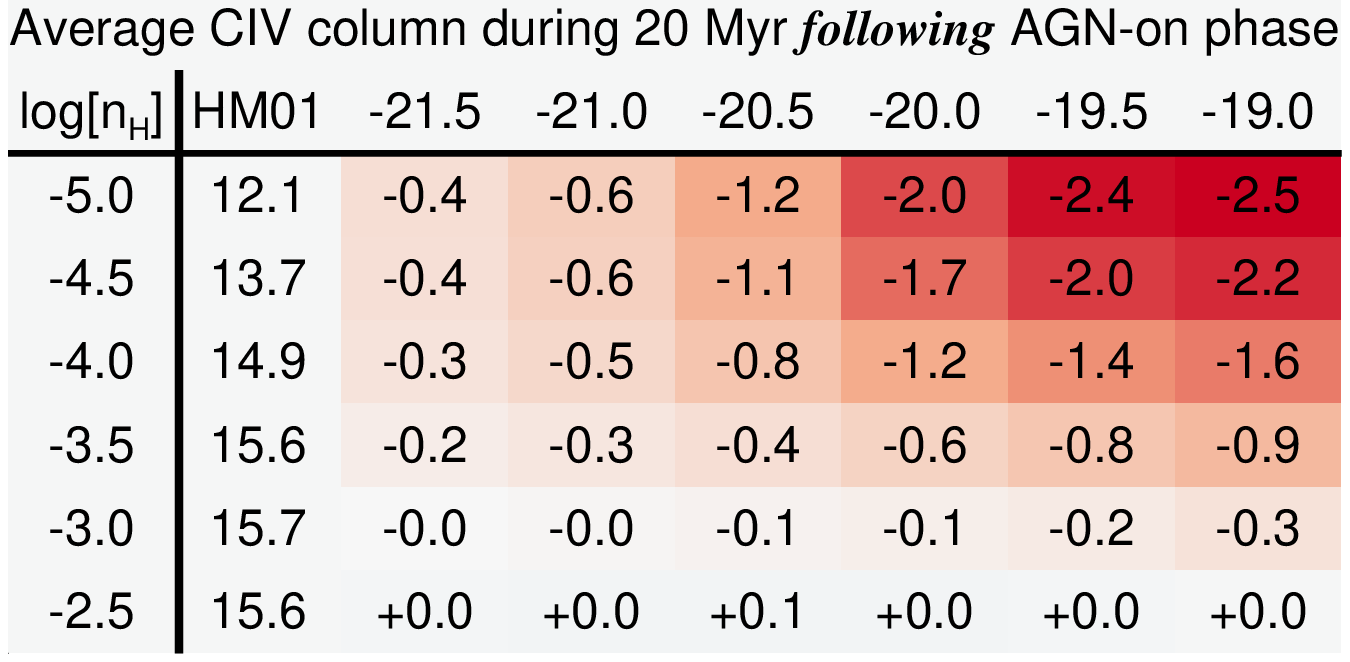}}
\subfigure{\setlength{\epsfxsize}{0.45\textwidth}\epsfbox{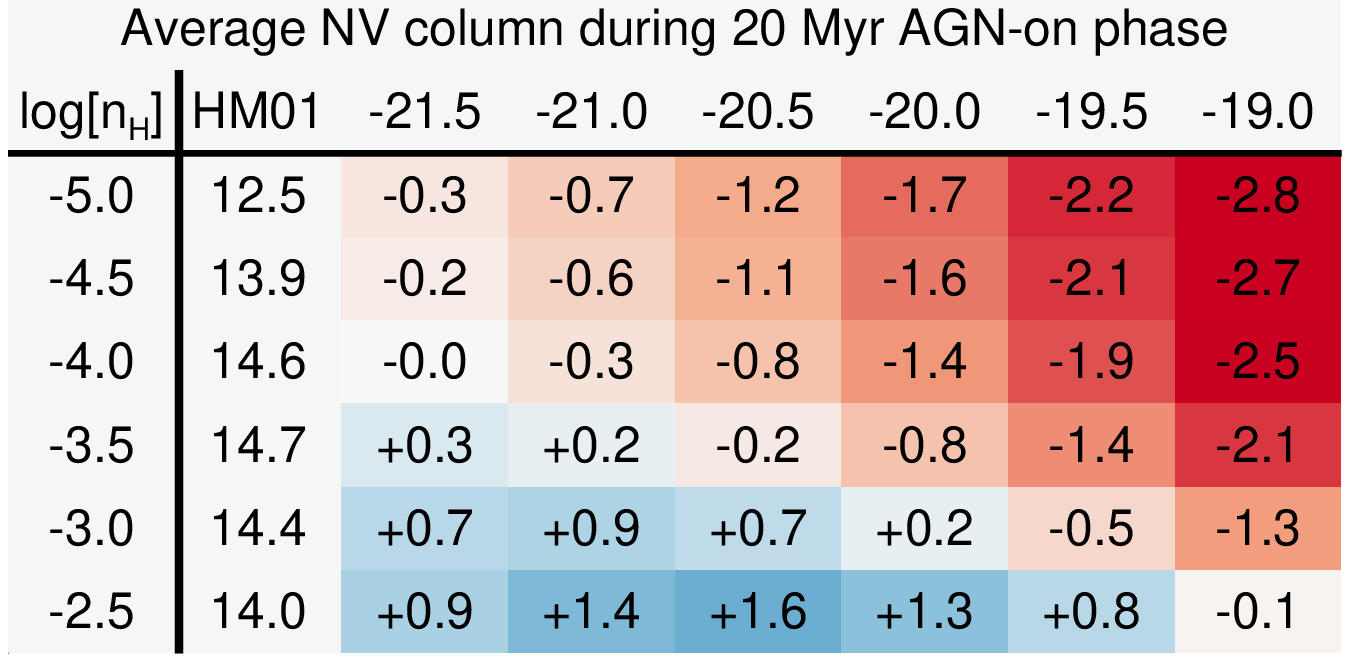}}
\subfigure{\setlength{\epsfxsize}{0.45\textwidth}\epsfbox{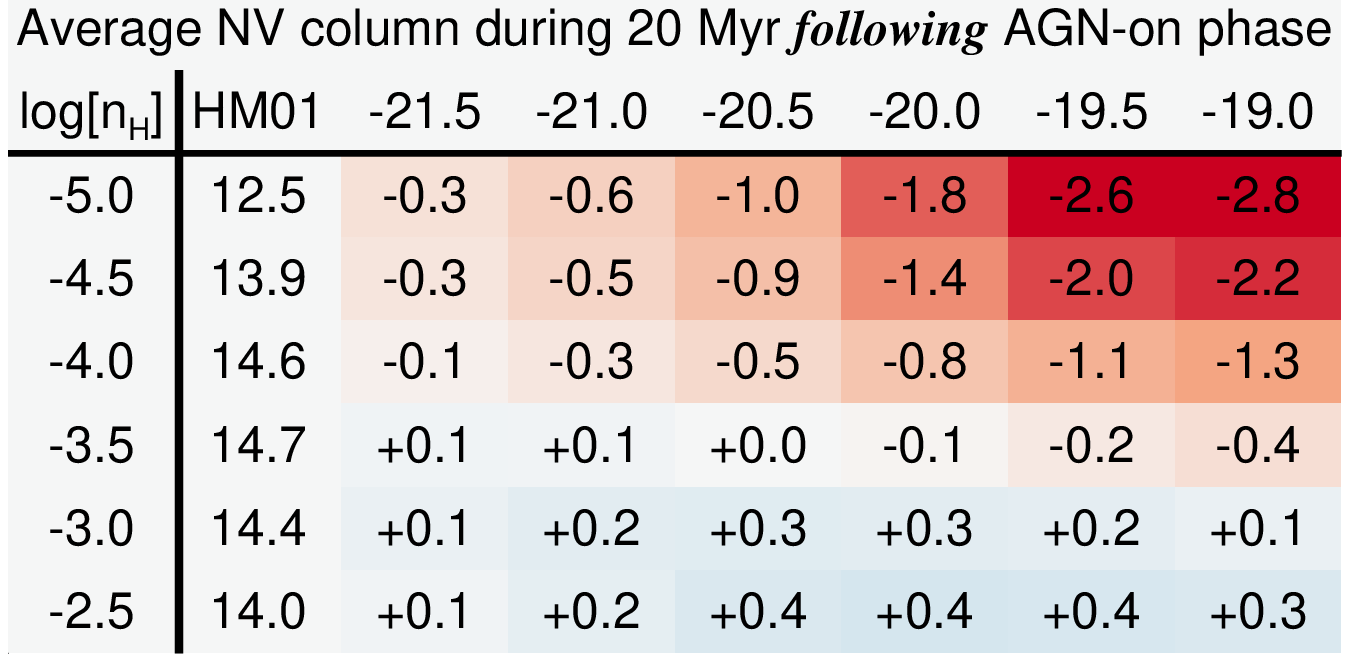}}
\subfigure{\setlength{\epsfxsize}{0.45\textwidth}\epsfbox{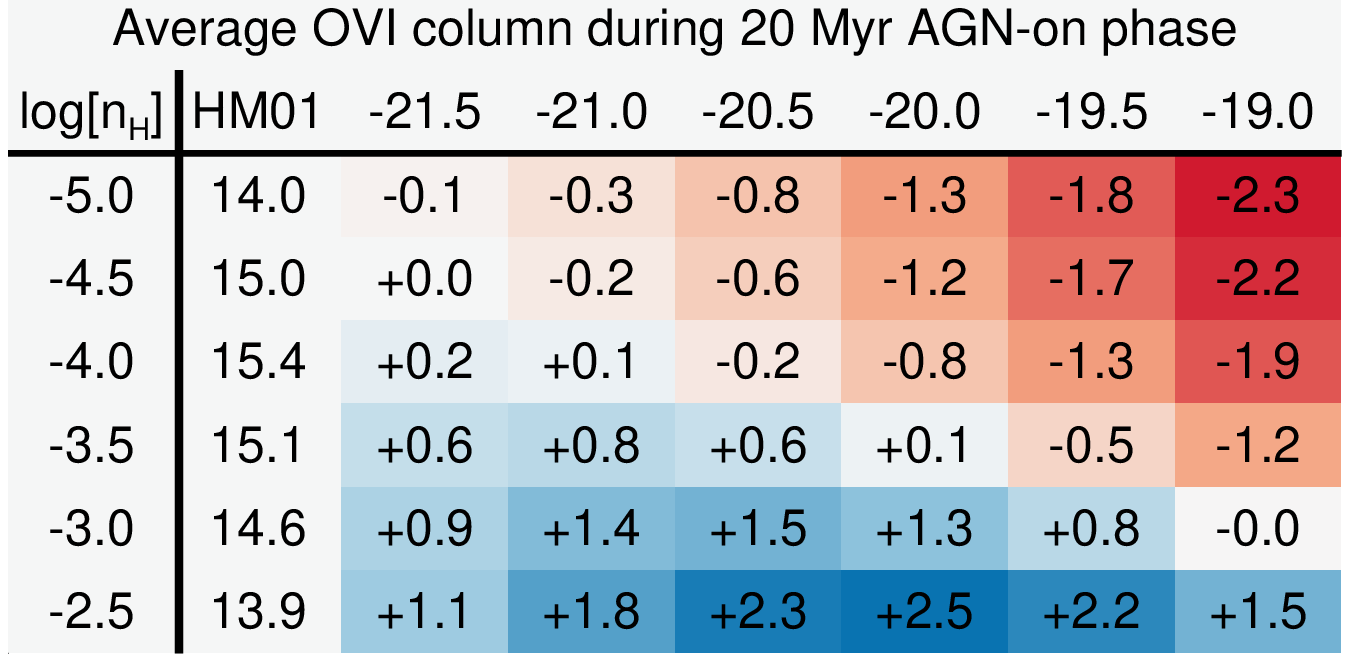}}
\subfigure{\setlength{\epsfxsize}{0.45\textwidth}\epsfbox{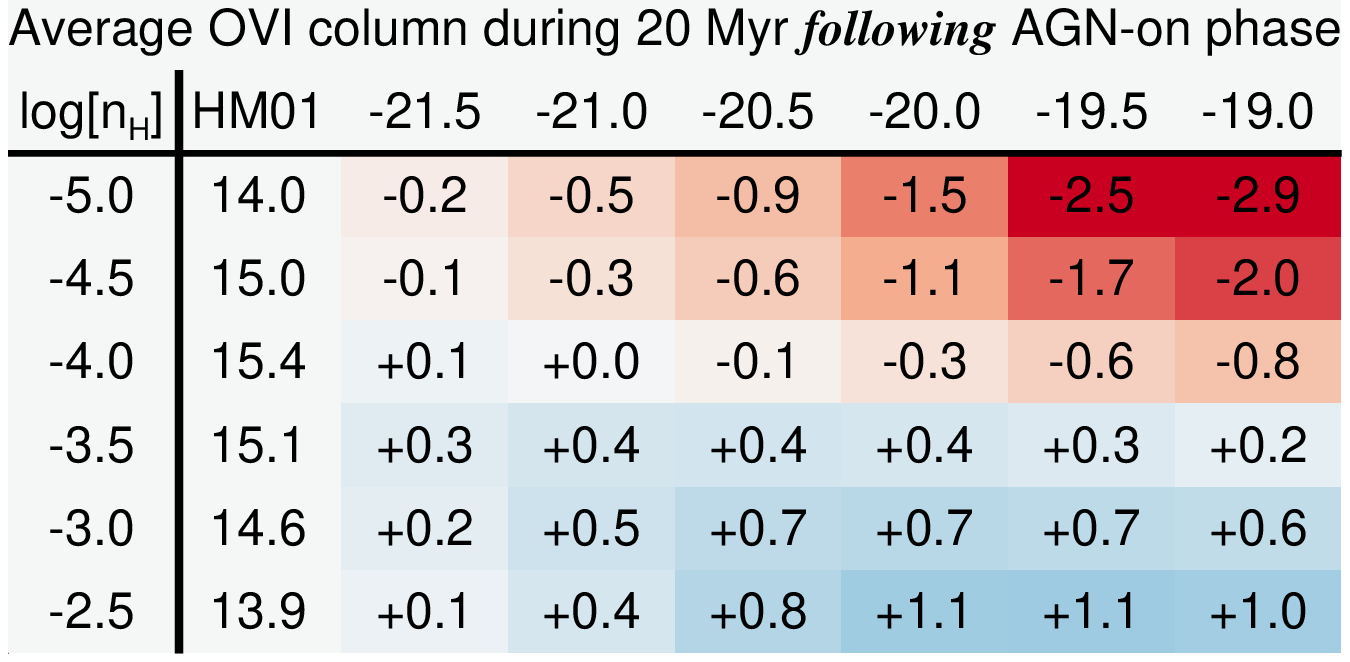}}
\caption[]{Time-averaged column densities of $\HI$ (top panels),
  $\CIV$ (2nd panels), $\NV$ (3rd panels) and $\OVI$ (bottom panels)
  for a $z=2.5$ AGN-on phase lasting 20 Myr (left panels) and the
  fossil effect of the 20 Myr immediately after the AGN turns off
  (right panels).  The 2nd column lists log[$N$/$\cms$] for the normal
  HM01 EGB at $z=0.25$, assuming ionisation equilibrium and solar
  metallicity, for 5 different densities (listed as log[($\cmc$)] in
  1st column).  The remaining columns indicate the effect of the AGN
  intensities (listed in the 2nd row in log[$J_{\rm LL,equ}$] units,
  where log[$J_{\rm LL,equ}$]$=-21.3$ at $z=2.5$ for the HM01 EGB)
  shown by the increase or decrease in log[$N$] relative to the HM01
  log[$N$].  Cell colour indicates whether the AGN increased (bluer)
  or decreased (redder) the column density.  $\HI$ is heavily affected
  during the AGN-on phase, but returns to nearly the original column
  afterwards.  $\CIV$, $\NV$, and $\OVI$ are also strongly affected
  during the AGN-on phase, and show time-averaged columns over the
  next 20 Myr that remain out of equilibrium.  Solar metallicity is
  assumed, but the metal columns scale with metallicity, except for
  the minor effect of AGN photo-heating raising the temperature
  slightly.  }
\label{fig:z25_QSO_after}
\end{figure*}

We turn on our AGN template spectrum for 20 Myr and find the average
change in column density owing to the enhanced field.  The left tables
in Figure \ref{fig:z25_QSO_after} show, from top to bottom, the
average enhancement in $\HI$, $\CIV$, $\NV$, and $\OVI$ of the
time-averaged column densities during the 20 Myr QSO-on phase for a
grid of densities (log[$\nh/\cmc$]) and fluxes (log[$J_{\rm
    LL,equ}/(\Jnuunits)$]).  The second column in each table shows
equilibrium column densities (log[$N/\cms$]) using the HM01 $z=2.5$
EGB, absorber lengths according to \citet{sch01}, and solar
metallicity.  As with all cases in this paper, we assume the gas
initially rests at the temperatures where photo-heating balances
cooling for the HM01 field, which fall from $10^{4.68}\rightarrow
10^{3.95}$ K as density increases from $\nh=10^{-5}\rightarrow
10^{-2.5} \cmc$ at $z=2.5$.  The grid cells to the right of the HM01
column indicate the column density change $\delta$log[$N$] (in dex)
relative to the HM01 equilibrium, colour-coded to be redder if the QSO
decreases the column and bluer if it is increased.

The behaviour of $\HI$ is relatively easy to understand.  As the AGN
turns on, $N_{\rm HI}$ equilibrates to a column density that is
inversely proportional to the intensity of the new radiation field.
We define the AGN enhancement factor as
\begin{equation}
g_{\rm AGN} = \frac{J_{\rm EGB}+J_{\rm AGN,equ}}{J_{\rm EGB}}.
\end{equation}
\noindent With the $z=2.5$ HM01 EGB having log[$J_{\rm LL}$]=$-21.3$
combined with the weakest AGN field log[$J_{\rm LL,equ}$]=$-21.5$,
$\HI$ is reduced by log[$g_{\rm AGN}$]$=0.2$ dex.  The six AGN field
enhancements we list range from $g_{\rm AGN}$=1.6-200, which decrease
$N_{\rm HI}$ in the same proportion modulo the effect of photo-heating
raising the temperature of the gas and thus further reducing $\HI$.
Photo-heating by the strongest AGN reduces $\HI$ by an additional
factor of 2-3 in $\nh\ga 10^{-3} \cmc$ gas, owing to the ejection of
energetic metal electrons at X-ray energies.

The behaviour of the metal lines is not as simple as that of $\HI$.
As we saw in the previous subsection, AGN almost always ionise $\CIV$
to higher states and thus reduce their columns, while enhancing $\OVI$
at higher $\nh$ and reducing it at lower $\nh$, depending on the field
strength.  We again assumed solar metallicity, but the listed
$\delta$log[$N$] values are applicable to other metallicities modulo
the effect of increased photo-heating at higher metallicity.  The same
trends are observed by \citet{gon08}: increased $\OVI$ and decreased
$\CIV$.  Their $g_{\rm AGN}$ ranged from $10-200$ and they derived
densities of $\nh=10^{-4.1}-10^{-2.7} \cmc$, which would correspond to
the bottom four rows and right-most four columns in the left tables of
Figure \ref{fig:z25_QSO_after}.  One difference is that they use a
single power law of $\nu^{-1.8}$ in their UV EGB, while we are using
the HM01 field enhanced with a $\nu^{-1.57}$ AGN spectrum plus an
X-ray component.  This allows our field to ionise $\OVI$ to $\OVII$
and above, which explains why we find more reduction in $N_{\rm OVI}$
for the stronger fields: $\OVI$ is ionised to higher states much more
easily.

An additional point about the TPE during the AGN-on phase is that
$\OVI$ increases significantly more than $\HI$ decreases at higher
densities and lower field strengths (see blue cells in bottom left
panel of Figure \ref{fig:z25_QSO_after}).  We may thus expect
significantly enhanced $\OVI$ with weaker than usual $\HI$ toward the
edge of a proximity zone, which may approach $\sim 10$ proper Mpc in
size for the brightest QSOs.  In fact, the large enhancement in
proximate $\OVI$ absorbers observed by \citet{tri08} within $1500$
$\kms$ of the brightest QSOs at lower redshifts (or $\sim17$ Mpc at
$z=0.5$), combined with their lower observed $N_{\rm HI}/N_{\rm OVI}$
ratios agrees with this prediction.  We put forth a model that these
are not absorbers directly associated with the QSOs, but instead
intervening IGM absorbers on the outskirts of the proximity zone with
a $g_{\rm AGN}$ of a few an perhaps even less than two.  This of
course assumes that the AGN has a harder EUV spectrum than the EGB.

A similar point can be made that $\NV$ could be enhanced in AGN
proximity zones.  Only cells with $\nh\geq 10^{-3.0} \cmc$ and weaker
AGN fields show noticeably enhanced $\NV$ when the AGN is on, but
these cells may be well-represented in QSO proximity zones.  Consider
that while $N_{\NV}$ for the equilibrium HM01 field peaks at
$\nh=10^{-3.5} \cmc$ (assuming $\Zsolar$ as in Figure
\ref{fig:z25_QSO_after}), the highest observed $\NV$ column may
correspond to $\nh \ga 10^{-3.0}$ given that metallicity likely
correlates with density.  Factoring in that most of the proximate
volume and path length corresponds to the weaker fields, the strongest
observable $\NV$ columns could correspond to the outskirts of
proximity zones.  Since nitrogen is less abundant than other metals
and $\NV$ is much rarer than $\CIV$ and $\OVI$, this AGN enhancement
could boost normally unobservable or hard-to-detect $\NV$ columns by
$\sim 5-10\times$ into obvious detections.

\subsubsection{QSO proximity zone fossils}

We now cross over to the fossil phase by considering the time-averaged
columns of $\HI$ and metals during the first 20 Myr after the QSO
turns off.  This timescale is chosen to demonstrate that significant
fossil effects can persist on a similar timescale as $\tAGN$.  We show
the deviations in columns on the right side of Figure
\ref{fig:z25_QSO_after}.  Listed is the time-averaged column deviation
over this time span (note that the actual deviation will likely be
larger or smaller at a specific fossil zone time).

The $\CIV$ column is almost always reduced, but not to the same extent
as during the QSO-on phase.  Nonetheless, if a significant fraction of
intervening $\CIV$ absorbers reside in fossil zones, then models that
try to fit these assuming a uniform EGB like HM01
\citep[e.g.][]{sch03, sim04, opp06, tes11} would underestimate the
amount of carbon.  Hence, the true global ionisation correction of
$\CIV$ would be larger, and the inferred IGM metallicity would be too
low.  The opposite would be true for $\OVI$ at higher densities, but
it would have the same trend as $\CIV$ if it mostly arises from lower
density gas.  Meanwhile, $\HI$ would remain nearly unaffected except
for the slightly reduced columns owing to photo-heating of metals.
The resulting absorption line systems will have ratios
(e.g.\ $\OVI/\HI$ and $\CIV/\HI$) that would differ significantly from
those predicted by equilibrium models with a normal EGB or in a
proximity zone.

\subsubsection{How affected are metal lines by fluctuating AGN?} \label{sec:volume}

We now present an estimate of the fraction of the volume of the
Universe that may be affected by transverse AGN proximity and fossil
zones at high-$z$.  We estimate the luminosity function
($\Phi(L_{\nu}$, \#/Mpc$^{-3}$/log($L_{\nu}$)) of AGN at 912\AA~using
the analytic fitting formulae of \citet{hopk07} at $z=2.5$\footnote{We
  used the script that was available on the website listed by this
  publication.}.  For a given isolated AGN, the proper radius within
which the AGN flux is at least $g_{\rm AGN}-1$ times $J_{\nu}$ is
given by $R= \sqrt{L_{\nu}/((4\pi)^2 (g_{\rm AGN}-1) J_{\nu}})$, where
$L_{\nu}$ is the AGN luminosity at $\nu$ in ergs s$^{-1}$ Hz$^{-1}$.
Hence, the fractional volume, $V_{\rm AGN}$, exposed to at least
$g_{\rm AGN}-1$ is
\begin{equation}
V_{\rm AGN} = \int{\phi(L_{\nu}) \frac{4\pi}{3} \left(\frac{L_{\nu}}{(4\pi)^2 (g_{\rm AGN}-1) J_{\nu}}\right)^{3/2} (1+z)^3 dL_{\nu} }
\end{equation}
\noindent when integrating over the luminosity function ($\nu$ is in
the rest frame).  The $(1+z)^3$ term is a result of $\Phi(L_{\nu})$
being in comoving Mpc and our proximity zone units being proper
units.  Using $J_{\nu}=10^{-21.3} \Jnuunits$ at $z=2.5$, the volume
ionised by AGN to at least the level of the HM01 background, $g_{\rm
  AGN}=2$, is 0.16\%, and the volume ionised to $g_{\rm AGN}=10$ is
0.005\%.  This calculation does not depend on how anisotropically AGN
emit, because the average subtended solid angle over which AGN of a
given luminosity emit is compensated by the number of all emitting AGN
for that luminosity.  The $V_{\rm AGN}$ fractions are much lower at
lower redshift, where AGN are less frequent and the Universe has
expanded.

While the fraction of volume with $g_{\rm AGN}\geq 2$ is small
compared to the volume filling factors of metals in simulations
calibrated to fit observed $\CIV$ statistics, which is of order 10\%
at $z\sim 2-3$ \citep{opp06,boo12}, there are reasons to argue that
AGN proximity and fossil zones could affect a significant fraction of
observed metal lines.  First, strong $\CIV$ systems ($N_{\rm CIV}\ga
10^{14} \cms$) are associated with Lyman-break galaxies (LBGs) at
small radii \citep[80-85 proper kpc][]{ade05,ste10}, and the co-moving
line density of such strong absorbers is consistent with being within
50 kpc of $\geq 0.5 L^*$ LBGs \citep{coo12}.  This implies that many
of the strongest metal-line absorbers occupy the small fraction of
volume with enhanced ionisation.  

More importantly, the multiplicative effects of AGN with short duty
cycles can significantly expand the volume affected by fossil zones.
Our lower redshift example in Figure \ref{fig:z09_life} shows that a
$d=4\%$ duty cycle leads to 3$\times$ greater $\OVI$ and 5$\times$
reduced $\CIV$ columns time-averaged once the AGN first turns on at
$t=0$.  This duty cycle is relevant since the observed fraction of
LBGs with active AGN is $\sim 3\%$ \citep{ste02}, which could reflect
their duty cycle.  Taking $1/d$ for $d=4\%$ means that the fossil
effect applies to a $25\times$ greater volume than the proximity zone
effect, and hence that 4\% of the total volume is affected by $g_{\rm
  AGN}\geq 2$.  This volume fraction is similar to that thought to be
required to account for the observed incidence of weak metal lines
\citep{opp06,boo12} suggesting that non-equilibrium AGN fossil zones
may be critically important.  Finally, metal lines can be enhanced if
$g_{\rm AGN}<2$, as is shown in Figure \ref{fig:z25_QSO_after} for
log[$J_{\rm LL,equ}$]=$-21.5$, or $g_{\rm AGN}=1.6$.  This leads to a
doubling of the volume affected by both proximity and fossil zones,
and $\CIV$ is reduced by $2-4\times$ in almost every case.  Again,
this arises because the AGN EUV spectrum is harder than that of the
EGB.

We conclude that a large fraction of metal lines could arise
from fossil zones at high-$z$, but precisely what fraction depends on many
factors.  If the AGN-on phase is long-lived and singular for most AGN
then the fossil effect would be limited.  However, if AGN are
fluctuating with short duty cycles in typical star-forming galaxy
hosts and have hard EUV slopes, then the fossil effect could be a
critical consideration for typical metal lines.  We may have to
re-evaluate the physical interpretation of $\CIV$ and $\OVI$ absorbers
at high-$z$, and the amount of underlying metals traced in diffuse
gas.

\subsection{Low-redshift Seyferts ionising their circumgalactic media} \label{sec:seyfert}\label{sec:app3}

\begin{figure}
\includegraphics[scale=0.7]{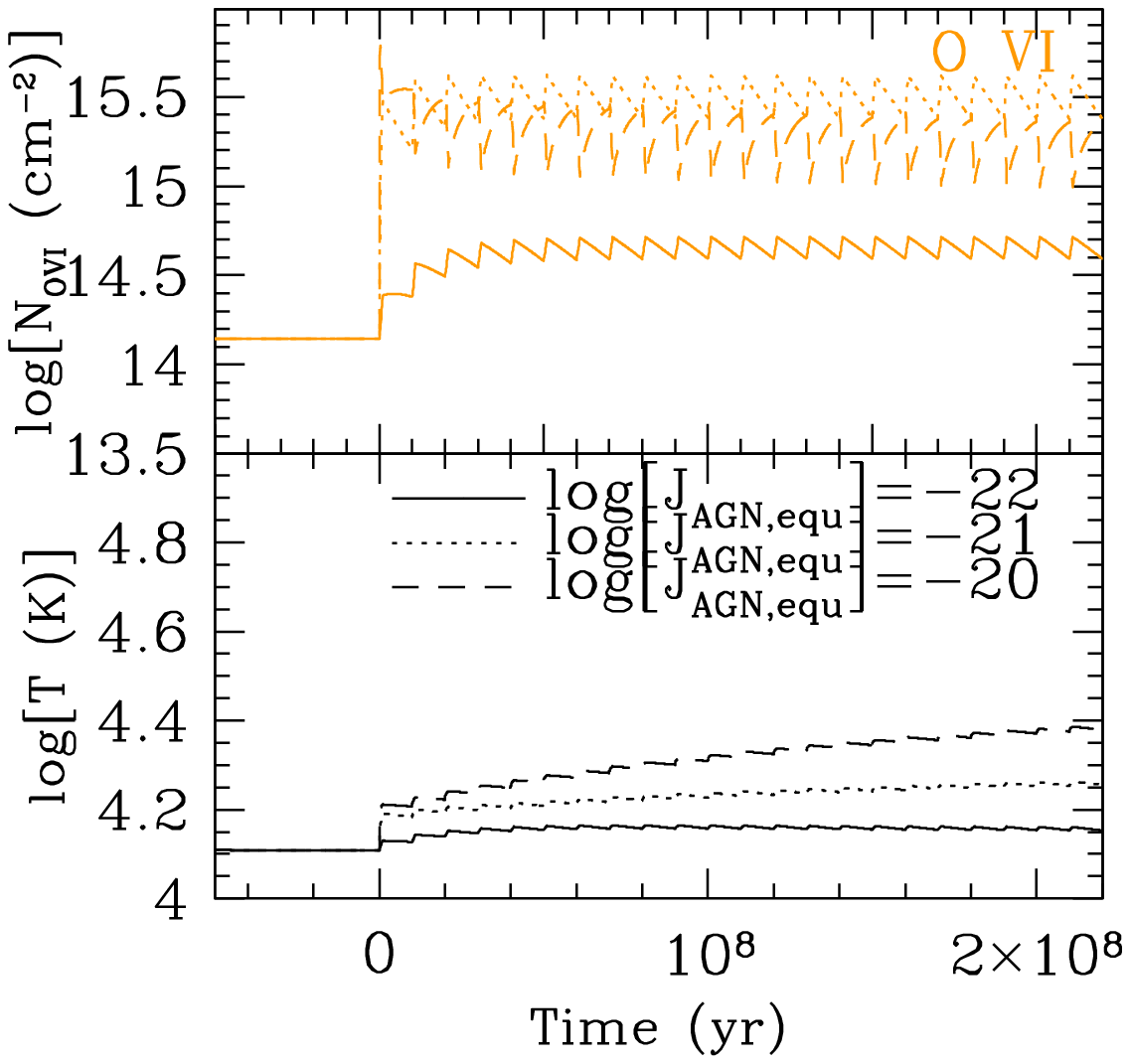}
\caption[]{The time evolution of $\OVI$ columns (top panel) and gas
  temperature (bottom panel) for a $\nh=10^{-4} \cmc$, $\l=74$ kpc
  solar metallicity absorber at $z=0.25$, where an AGN of the listed
  intensity turns on for 1 Myr every 10 Myr starting at $t=0$.  The
  periodic AGN activity significantly enhances $\OVI$ columns and
  doubles the gas temperature over 200 Myr in the most extreme case
  via photo-heating of metals.  This case could be applicable for
  metals around normal looking galaxies if AGN duty cycles are short
  and have frequencies comparable to the $\OVI$ recombination
  timescale.}
\label{fig:z025_Seyferthist}
\end{figure}

Our last case study are low-redshift metal absorbers associated with
weaker AGN, such as Seyferts.  We consider an admittedly idealised
situation of a 10\% duty cycle where the AGN turns on for 1 Myr out of
every 10 Myr at $z=0.25$.  We plot the time history of $\OVI$ in
Figure \ref{fig:z025_Seyferthist} for three different field
enhancements, to show how the continuous pumping resulting from a
short duty cycle can even significantly enhance the ionisation around
weak AGNs.  We again assume $\nh=10^{-4} \cmc$ solar-enriched gas at
the equilibrium temperature of $T=10^{4.1}$ K.

This situation may be applicable to Markarian galaxies in the local
Universe.  These are among the biggest, brightest spirals, and have an
unusually blue central core that usually requires non-thermal
emission, which often indicates AGN activity.  About 10\% of
Markarians are Seyferts \citep{wee77}, which motivates our choice of a
10\% duty cycle.  Of course, not all Markarians are Seyferts with a
short duty cycle, but we consider the possibility that $\sim$10\% of
Markarians were more active AGNs in the recent past (i.e.\ in the last
$\sim 10$ Myr).  The three cases plotted in Figure
\ref{fig:z025_Seyferthist}, $J_{\rm LL,equ}= 10^{-22.0}$,
$10^{-21.0}$, and $10^{-20.0} \Jnuunits$ correspond to soft X-ray
luminosities at 100 proper kpc of $L_{0.5-2.0}=10^{42.1}$,
$10^{43.1}$, and $10^{44.1} \ergs$ AGN, respectively, using our AGN
template spectrum.  These luminosities are in the range of Seyfert
X-ray luminosities observed by \citet{rus96}.  The bottom panel of
Figure \ref{fig:z025_Seyferthist} shows that the temperature doubles
to $10^{4.4}$ K in the strongest case.

\begin{figure}
\subfigure{\setlength{\epsfxsize}{0.45\textwidth}\epsfbox{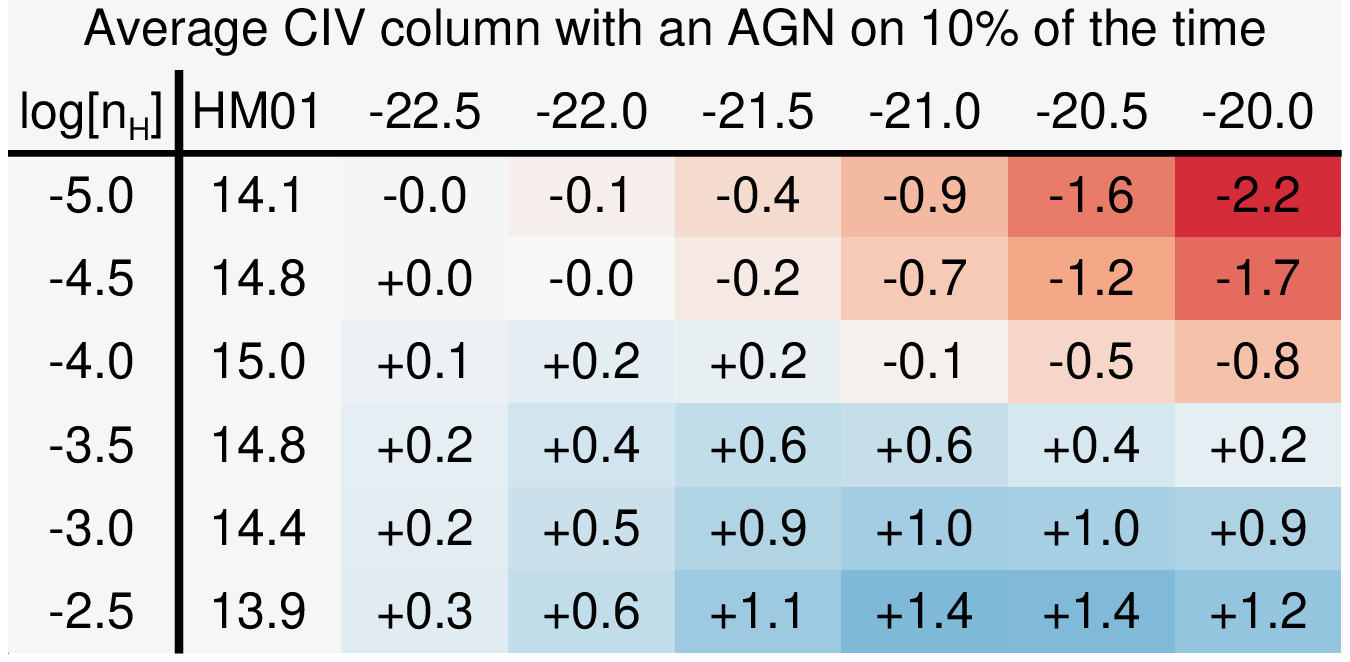}}
\subfigure{\setlength{\epsfxsize}{0.45\textwidth}\epsfbox{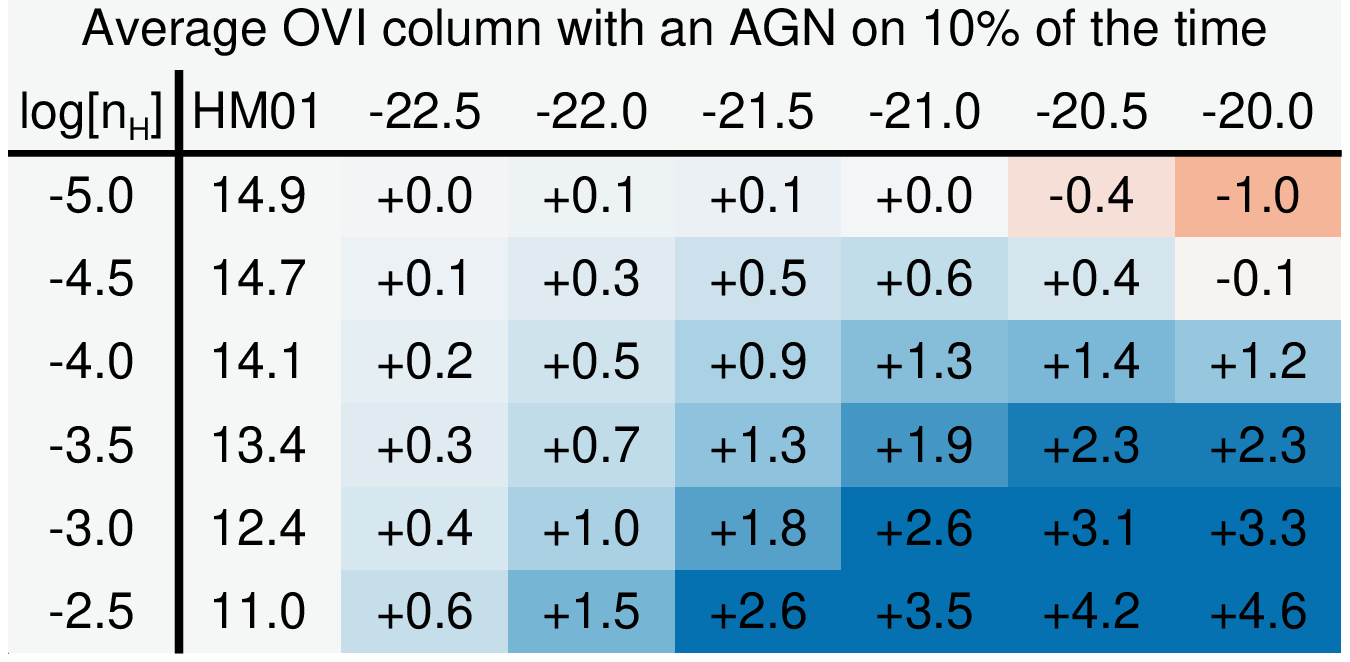}}
\subfigure{\setlength{\epsfxsize}{0.45\textwidth}\epsfbox{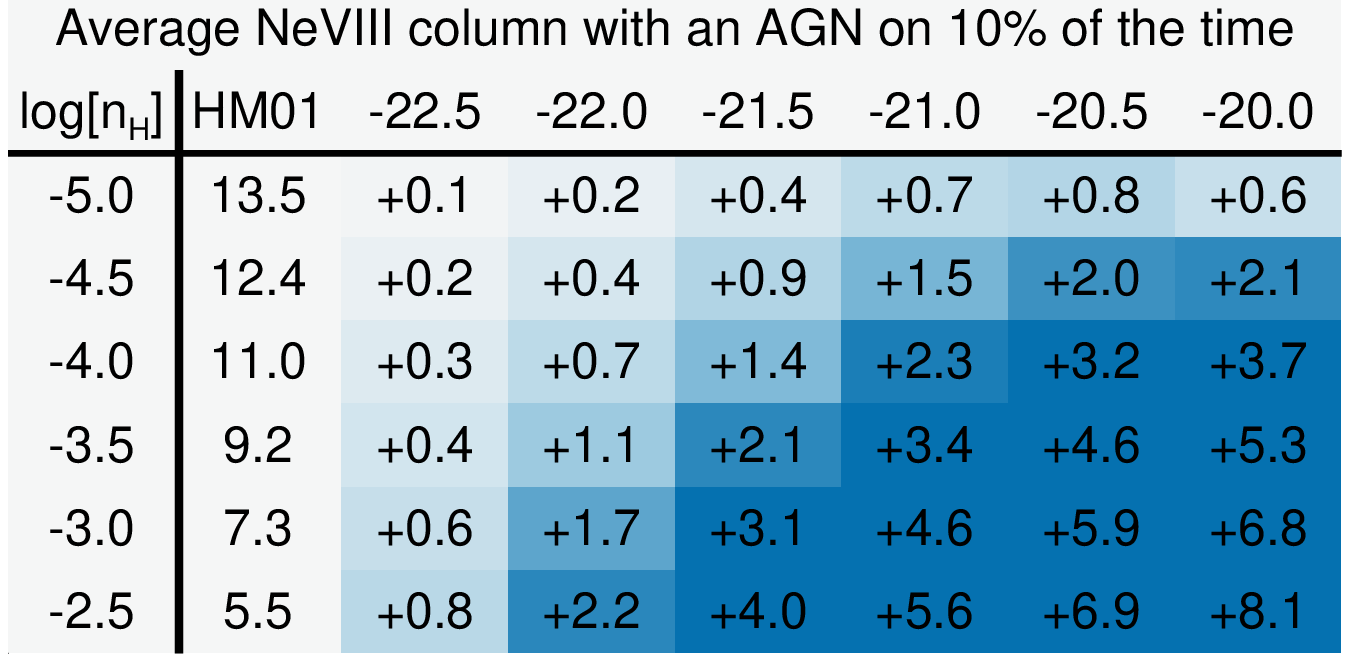}}
\caption[]{Time-averaged column densities of $\CIV$ (top panel),
  $\OVI$ (middle panel), and $\NeVIII$ (lower panel) assuming that a
  $z=0.25$ AGN turns on for 1 Myr every 10 Myr.  The 2nd column lists
  log[$N$ ($\cms$)] for the normal HM01 field at $z=0.25$, assuming
  solar metallicity, for 5 different $\nh$ densities (listed as
  log[($\cmc$)] in the 1st column).  The rest of the columns indicate
  the time averaged effect of the AGN intensities (listed in the 2nd
  row in log[$J_{\rm LL,equ}$] where log[$J_{\rm LL,equ}$]$=-22.2$ at
  $z=0.25$ for the HM01 EGB) shown by the relative increase or decrease in
  log[$N$] relative to the HM01 log[$N$].  Cell colour indicates
  whether the AGN increases (blue) or decreases (red) the column
  density.  Despite the AGN being on only 10\% of the time, metal
  columns can be affected, mostly in a positive direction.  Solar
  metallicity is assumed, but the metal columns scale with
  metallicity, except for the minor effect of AGN photo-heating
  raising the temperature slightly.  }
\label{fig:z025_Seyferttable}
\end{figure}

We tabulate the time-averaged enhancements of $\CIV$, $\OVI$, and
$\NeVIII$ over 200 Myr, or the first 20 cycles, for a grid of
densities (log[$\nh$/$\cmc$]) and AGN fluxes (log[$J_{\rm
    LL,equ}$/($\Jnuunits$)]) in Figure \ref{fig:z025_Seyferttable}.
Similar to Figure \ref{fig:z25_QSO_after}, the second column in each
table shows column densities (log[$N$/$\cms$]) using the HM01 EGB
(here at $z=0.25$), absorber lengths according to \citet{sch01},
$T=10^{3.80-4.40}$ K before the AGN turns on, and solar metallicity.
The grid cells to the right indicate the column density change
$\delta$log[$N$] relative to the HM01 equilibrium, colour-coded to be
redder if the column density goes down due to the AGN and bluer if it
goes up.  The $\nh$ densities correspond to overdensities ranging from
27 to 8500.  While we show solar metallicity, the change is similar
for lower metallicities.

The main point of this diagram is to show that even for a 10\% AGN
duty cycle, the time-averaged columns are significantly changed,
mostly in a positive direction for these higher metal ions.  While
there has never been a survey designed to probe the halos of galaxies
showing signs of on-going or recent AGN activity in the low-redshift
Universe, it may be worth exploring in light of the COS observations
of \citet{tum11} revealing very strong $\OVI$ around normal spiral,
star-forming galaxies at $z\sim 0.2$.  Markarian galaxies, for
example, may show even stronger $\OVI$ if this gas originates from a
photo-ionised component at $T<10^{5}$ K.

\section{Summary} \label{sec:summary}

The assumption of a uniform extra-galactic background (EGB) breaks
down in an AGN proximity zone, which is over-ionised relative to the
rest of the IGM.  $\HI$ columns are reduced and metals are ionised to
higher states owing to increased photo-ionisation.  We demonstrate
here that once the AGN turns off, metals residing in ``fossil'' AGN
proximity zones may remain over-ionised for timescales exceeding the
typical AGN-on phase.  Using the non-equilibrium code introduced in
\citet{opp13a}, we follow metal-enriched gases initially ionised by
the EGB \citep{haa01} under the effect of increased ionisation from a
local AGN.  After the AGN turns off, metal ions in fossil proximity
zones often require many Myr to return to the equilibrium values,
owing to the significant recombination times of metal ion species.  We
demonstrate how this non-equilibrium fossil effect alters
observational metal-line diagnostics used to constrain the physical
state of intergalactic and circum-galactic gas.  In particular, we
show that commonly observed Lithium-like ions (i.e.\ $\CIV$, $\NV$,
$\OVI$, $\NeVIII$) can be heavily affected at a range of redshifts
from the present-day to z$\geq 3$.  This behaviour is in contrast to
$\HI$, which re-equilibrates comparatively rapidly once an AGN turns
off, owing to its equilibrium state having very low neutral fractions
in the IGM.

We consider metal-enriched gas with densities $\nh=10^{-5} - 10^{-2.5}
\cmc$ at $z=2.5, 0.9$ and 0.25, which corresponds to diffuse IGM and
CGM gas, that is initially in thermal equilibrium ($T\sim
10^{4}-10^{4.5}$ K).  Our parameter choices are motivated by
hydrodynamic simulations, which show a significant fraction of diffuse
metals in thermal equilibrium, where cooling balances photo-heating,
owing to short cooling times from higher temperatures ($T\sim
10^5-10^6$ K).  The gas begins in ionisation equilibrium with the HM01
EGB, and becomes irradiated by an AGN template spectrum turned on for
1 to 20 Myr corresponding to an ionisation increase of between
$1.6-500\times$ the EGB radiation field strength at the Lyman limit.

Our fiducial case corresponds to a local AGN turning on at $z=0.9$, a
redshift for which a wide range of metal lines are observable with
current facilities (e.g.\ COS on HST) and which corresponds to the
\citet{tri11} observations of very strong $\NeVIII$ absorption in a
system associated with a galaxy exhibiting post-starburst and possible
AGN signatures.  We show that for reasonable AGN energies, $\NeVIII$
columns are increased by a factor of $\sim 100$ during a 20 Myr AGN-on
phase and also during the 15 Myr in the following fossil phase, and
remain $> 10\times$ stronger for 50 Myr after the AGN turns off.  We
show for a variety of gas densities, AGN strengths, and AGN lifetimes
that $\NeVIII$ columns of $10^{14}-10^{15} \cms$ are possible for
significant timescales during the fossil phase for solar abundances.
We directly model the components of the $z=0.927$ PG1206+459 absorber
\citep{tri11}, finding a single phase model that successfully fits
$\NeVIII$, $\MgII$, $\HI$, and other metal columns assuming
$\nh=10^{-3} \cmc$, $T\sim 10^4$ K gas in a fossil zone several Myr
after a strong QSO turned off.  Our application to this absorption
system demonstrates that a whole new class of non-equilibrium
solutions are available for metal-enriched diffuse gas in the presence
of evolving ionisation fields.

We also model AGN ionisation at $z\sim 2.5$, including the transverse
proximity effect for a foreground QSO zone in a QSO pair, where
observed $\CIV/\HI$ and $\OVI/\HI$ line ratios indicate enhanced
ionisation over the normal EGB \citep{gon08}.  We first show that
active proximity zones can exhibit these unique ratios, and then
consider how the non-equilibrium fossil effect can alter $\CIV$ and
$\OVI$ columns for timescales longer than the AGN lifetime.  Unique
line ratios are possible where $\HI$ has rapidly re-equilibrated but
the metal lines are still recombining.  

We argue that it is plausible that fossil zones could affect many more
absorbers in the high-$z$ Universe than AGN proximity zones if AGNs
fluctuate with short duty cycles, have hard extreme UV slopes, and are
associated with typical star-forming galaxies, which prolifically
enrich the IGM.  A large fraction of intervening metal absorbers in
typical QSO spectra may well be affected by the fossil effect, meaning
that ionisation corrections assuming a uniform EGB would be wrong.
$\CIV$ columns are almost always reduced in fossils meaning that the
true diffuse gas metallicity becomes higher than when one assumes
equilibrium with the EGB field.

Finally, we consider lower redshift AGN ionising their surrounding CGM
with a 10\% duty cycle where the AGN is on for 1 out of every 10 Myr.
Despite this short duty cycle, high ions ($\OVI$ and $\NeVIII$) are
significantly enhanced, often by more than 1 dex, and temperatures are
increased by a factor of two for stronger fields.  The key to these
increases of metal columns and temperature are that the
photo-ionisation/photo-heating timescales are short or comparable to
the AGN lifetime, while the recombination/cooling timescales are long
or comparable to the duty cycle interval.  The result is intriguing in
that surprisingly little energy input is necessary to greatly enhance
metal columns in the halo of fluctuating typical AGN.

These non-equilibrium fossil and proximity zone effects may complicate
the analysis of intergalactic metal lines that assume photo-ionisation
equilibrium with a uniform EGB.  For example, cosmological hydrodynamic
simulations using such assumptions underestimate the observed number
density of the highest equivalent width $\NeVIII$ \citep{opp12a,tep13}
and $\OVI$ absorbers \citep{opp09a,tep11}, unless a sub-grid turbulence
model is invoked post-run.  Proximity zone fossils could provide an
alternative model for such strong, highly ionised metal absorbers.

The fossil effect is unique in that $\HI$ re-equilibrates rapidly
while metal lines remain over-ionised leading to different ionisation
corrections to obtain metallicities.  This could be particularly
important during the quasar era, $z\sim 2 - 5$, for which we find that
the fraction of the volume occupied by fossil zones may be similar to
the fraction that is thought to be enriched with heavy elements.

\section*{Acknowledgements}

We are grateful for discussions with Sebastiano Cantalupo, Arlin
Crotts, Romeel Dav\'e, Kristian Finlator, Martin Haehnelt, Zoltan
Haiman, Phil Hopkins, Mike Shull, Chuck Steidel, John Stocke, Todd
Tripp, and Lisa Winter.  We thank Ali Rahmati for a thorough reading
of this manuscript, and the anonymous referee for a constructive
review.  This work benefited from financial support from the
Netherlands Organisation for Scientific Research (NWO) through VENI
and VIDI grants, from NOVA, from the European Research Council under
the European Union's Seventh Framework Programme (FP7/2007-2013) / ERC
Grant agreement 278594-GasAroundGalaxies and from the Marie Curie
Training Network CosmoComp (PITN-GA-2009-238356).  We are also
thankful for the hospitality provided by the University of Colorado,
Boulder where part of this work was completed.

\end{document}